\documentclass[reprint,amsmath,amssymb,aps,]{revtex4-2}
\usepackage{graphicx}
\usepackage{dcolumn}
\usepackage{braket}
\usepackage{bm}
\usepackage{subfigure}

\begin{document}
\preprint{APS/123-QED}
	
\title{Wide Sampling and Efficient Updating Monte Carlo Algorithms for Dimer Models}
\author{Hongxu Yao\textsuperscript{1}}\thanks{These authors contributed equally with Jiaze Li and Jintao hou. Corresponding auther: 18307110355@fudan.edu.cn}
\author{Jiaze Li\textsuperscript{1}}\thanks{These authors contributed equally.}
\author{Jintao Hou\textsuperscript{1}}\thanks{These authors contributed equally.}
\address{1,Department of Physics and State Key Laboratory of Surface Physics, Fudan University, Shanghai 200438, China}
\date{\today}

\begin{abstract}
	Quantum dimer model is a low-energy and efficient model to study quantum spin systems and strong-correlated physics. As a foreseeing step and without loss of generality, we study the classical dimers on square lattice by means of Monte Carlo method. For efficient states updating in dimer model, we introduce a highly-efficient loop updating algorithm directed by energy criterion called energy directed loop algorithm and improve the pocket algorithm to compare them with the traditional directed loop algorithm. By comparisons, our energy directed loop algorithm increases the convergent speed of Monte Carlo and shorten the auto-correlated time in classical hard-core dimer model. Both the improved pocket algorithm and energy path algorithm can be used in varietal dimer models and succeed in traversing the topological sections rapidly.

\end{abstract}

\maketitle
\section{Introduction}
	The quantum dimer model (QDM) is a low-energy and efficient model proposed by D.S. Rokhsar and S.A. Kivelson in 1988. It acts as an important way to understand the quantum spin liquid\cite{PhysRevLett.61.2376}. In QDM, two adjacent 1/2 spins will be bonded and two bonds form exponentially decaying short-range coupling called resonating valence bond (RVB) proposed by P.W. Anderson to explain the properties of high temperature superconductivity\cite{doi:10.1126/science.235.4793.1196,doi:10.1080/14786439808206568,BASKARAN1987973}. 	
	
	For representing the spin-spin correlation and simplifying RVBs in lattices, there is exactly one bond with the nearest point for every point in the lattice as a dimer and two parallel dimers in a plaquette can interact with each other forming RVBs as Fig.1(a). Due to the lack of efficient updating methods, researchers mostly use local approximation method at quantum critical points but some conclusions are still controversial\cite{PhysRevB.71.020401,PhysRevB.73.245105,PhysRevLett.100.037201}. Lately, researchers have found emergent topological phases and demonstrated complex quantum transition mechanism in QDMs as the development of computational power and efficient algorithms. By means of sweet cluster algorithm\cite{PhysRevB.99.165135}, in square lattice researchers confirm all phases are mixed phases in the left of Rokhsar-Kivelson point\cite{PhysRevB.103.094421}. In frustrated triangular lattice, they achieve a quantum transition from quantum spin liquid state to valence bond solid state with a $\sqrt{12}\times \sqrt{12}$ emergent order\cite{2021Topological}. 

	For deeper studying QDMs, we need more exact or numeric methods to overcome quantum many body questions. Classical dimers with hard-core condition is a special and reduced case without kinetic term of QDM Hamiltonian, which gives our chance to develop new algorithms. In our study, we reappear the traditional directed loop algorithm (DLA) in classical dimer model by means of Monte Carlo\cite{doi:10.1063/1.1632141,PhysRevE.71.036706} and introduce energy criterion to achieve more efficient update. The new algorithm is called energy directed loop algorithm (eDLA). Not only can eDLA form longer loops to update more dimers once, but it also provides larger sampling space in the whole ensemble. Longer loop means the decrease of convergent times in a Monte Carlo step and data correlation will be also lessened.

	Meanwhile, if we break hard-core condition in dimer model such as considering next nearest neighbor correlation between two spins\cite{2004Deconfinement} and so on, more unknown physics will emerge from them. However, we still have no more efficient algorithms to solve these questions. We improve pocket Monte Carlo algorithm\cite{PhysRevB.67.064503} called pocket edge algorithm (PEA) so that all three algorithms can adapt to different in varietal dimer models called softed dimer models (SDMs) generally. We compare them and try to derive the most efficient methods applying to solve dimer questions. 

\section{ENERGY DIRECTED LOOP ALGORITHM AND POCKET EDGE ALGORITHM FOR MONTE CARLO SIMULATION}

\subsection{Energy directed loop algorithm}
	Traditional DLA is introduced by A. W. Sandvik et al in $1/2$ antiferromagnetic Heisenberg model. Previous researchers improve it to dimer model as a universal algorithm. Its detailed process will be shown in Appendix A (see Fig.8).

	Recall that Monte Carlo simulation will be constructed under the fundamental of Metropolis sampling algorithm\cite{2008LNP...739.....F}, we decompose the whole updating algorithm into every steps of the generation of directed loops in our new algorithm now. Different from the previous method calculating the energy change between the initial state and the final state after a single update, we calculate the energy change accompanying the growth of loop.  

	The whole process is in the following:

	1). We choose a vertex as our starting point and annihilate the occupied dimer here. After that, we define the other point of the dimer as a walking seed. In Fig.1(a), we choose the blue point as the starting point and the red as our walking seed.

	2). For the point of walking seed, we have four directions to create new dimer as our path like the Fig.1(b). It means we can create new dimer in four directions around the walking seed. The directed loop will step into one of these four directions as red arrows showing.

	3). It is distinguishable to the regular directed loop, instead of just choosing one direction to walk randomly, we compute the energy change of every path in four choices of direction and judge each of them by a prior Metropolis sampling. For example, in Fig.1(c), we suppose the direction pointing to the top has been chosen. Thus, we compute the energy change if we create a new red dimer here. The red dimer can bond with the parallel dimer in the yellow dashed line frame. If the energy change in the frame is accepted, we can derive a path of the directed loop. 

	4). Then, we annihilate the cross-over dimer and repeat the above step until the energy directed loop comes back to the starting point and form a closed loop. In Fig.1(d), the directed loop repeats above operations. 

	If directed loops walk randomly, DLA either generates too long loop to be accepted by the system or it forms local small loops so that the system drops into sub-equilibrium states flipping over and over again among a series of energy-adjacent states. Both of two above cases are a great waste of computational power. Now, consider eDLA, the addition of energy criterion induces a more efficient walking in the updating process in Monte Carlo simulation. Real-time Metropolis sampling judging in every steps of the generation of the energy directed loop depends the loop length and the energy change matching with the homologous thermal states. 

	Especially, in two cases, energy directed loop matching to the real process can advance the simulation to a great extent. Firstly, in large scale cases, it is a difficult problem to handle large or miscellaneous lattice for traditional DLA such as  $L\times L > \approx100\times 100$ or on the kagome lattice\cite{PhysRevLett.89.137202} or Penrose tiling\cite{PhysRevX.10.011005}. If the scale is too large, we need to simulate hundreds of millions of times. At a finite temperature, the system needs too long time to relax constrained by the slow convergent speed and long auto-correlated time. Secondly, we can define topological sections in dimer lattice with periodic boundary condition\cite{PhysRevB.54.12938,yan2021improved}. Traditional DLA is a local algorithm and it is difficult to form global perfoliate loop traversing topological sections but eDLA can introduce global loop widening the effective sampling ensemble that could include more topological sections. More detailed data will be given in  Sec. III.
	\begin{figure}[htbp]
		\centering
		\subfigure[]{ \label{1a} \includegraphics[scale=0.25]{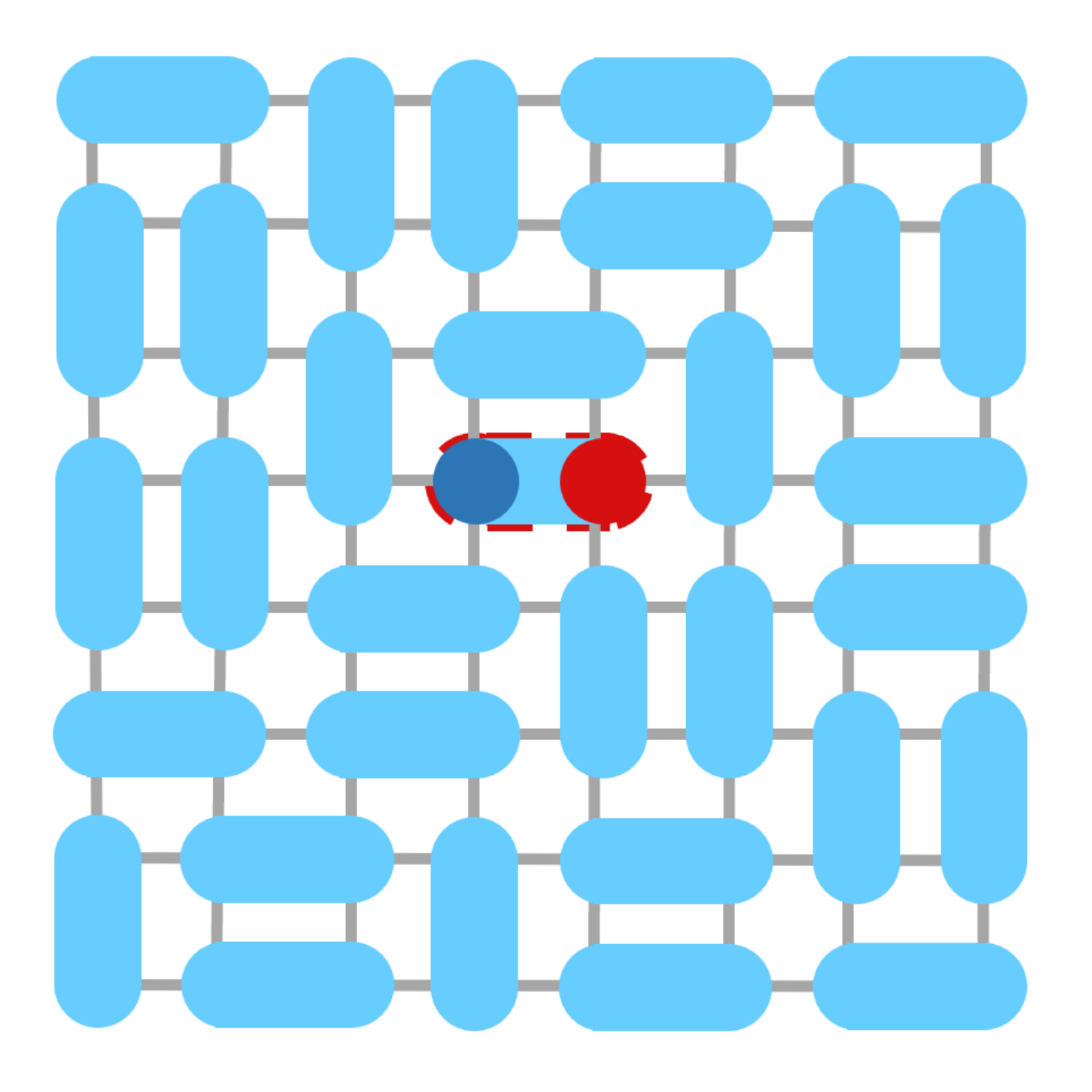}}
		\quad		
		\subfigure[]{ \label{1b} \includegraphics[scale=0.25]{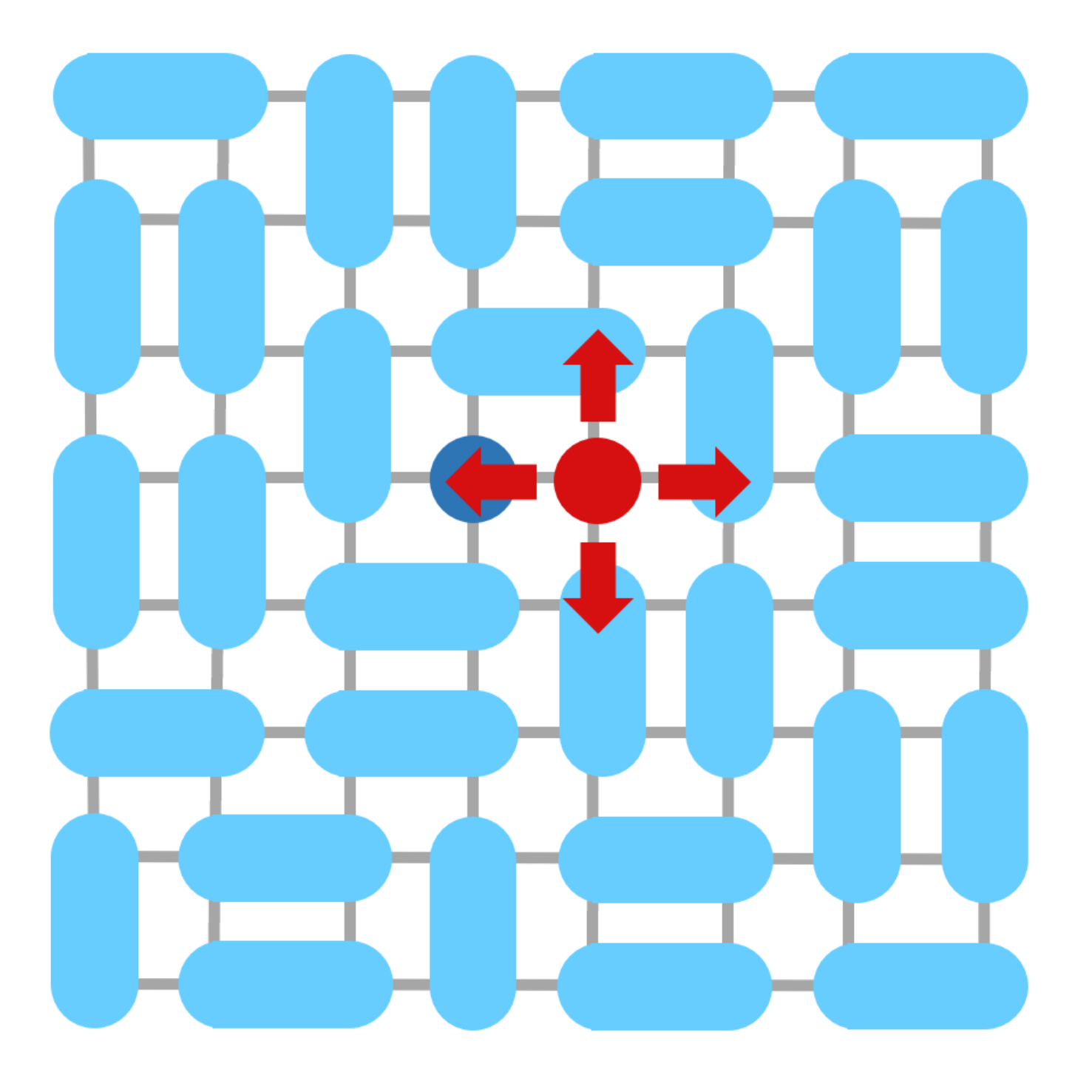}}
		\quad	
		\subfigure[]{ \label{1c} \includegraphics[scale=0.25]{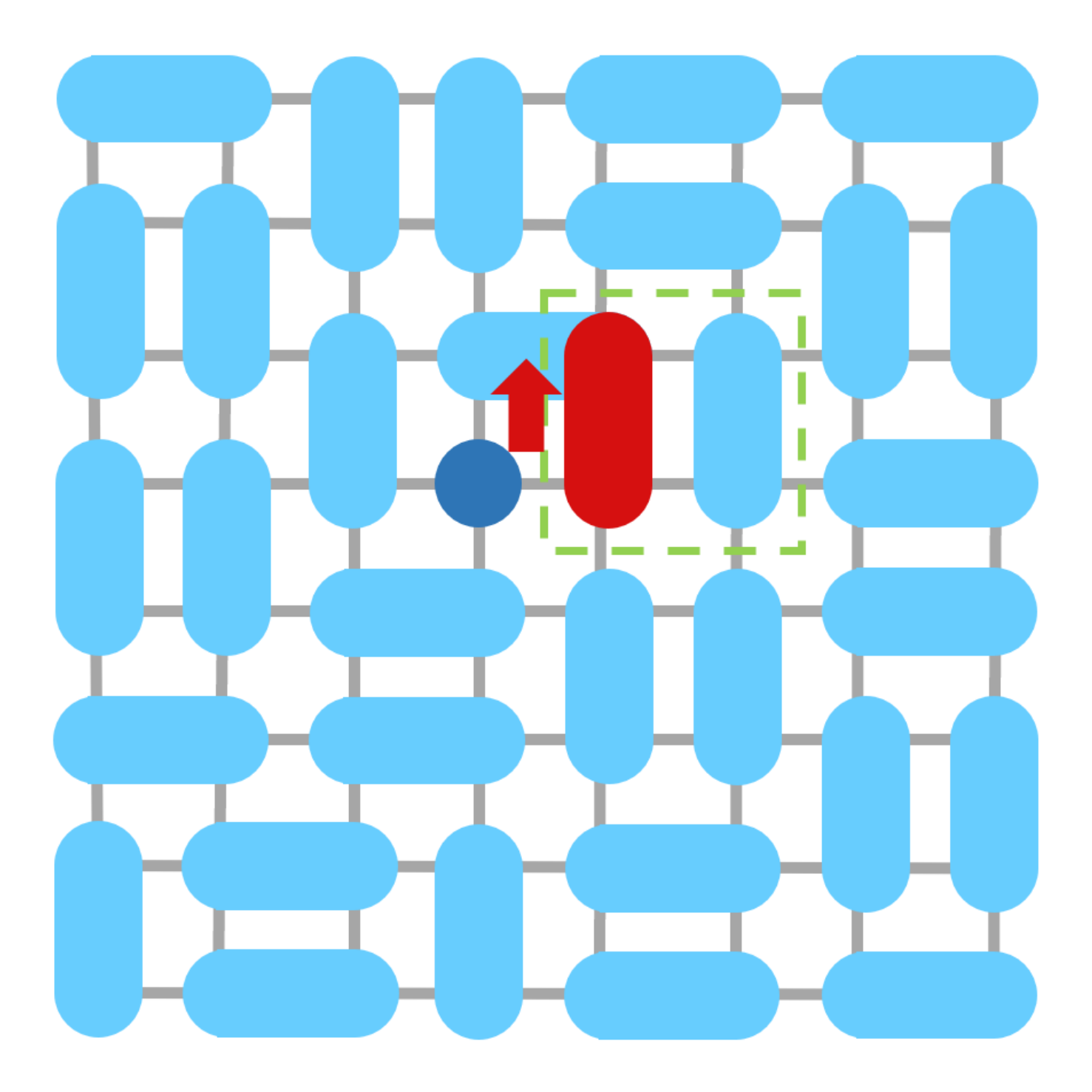}}
		\quad
		\subfigure[]{ \label{1d} \includegraphics[scale=0.25]{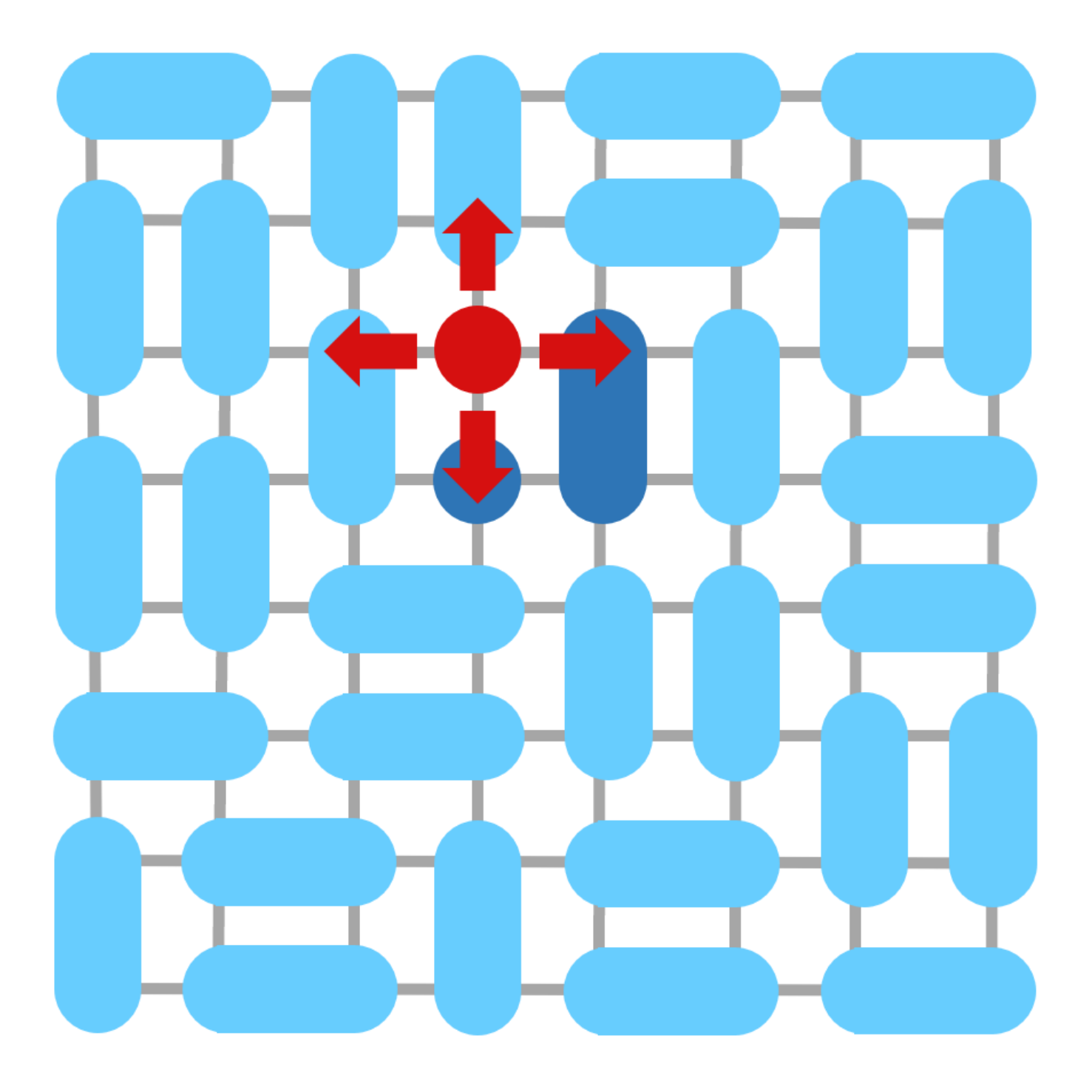}}
		\quad		
		\subfigure[]{ \label{1e} \includegraphics[scale=0.25]{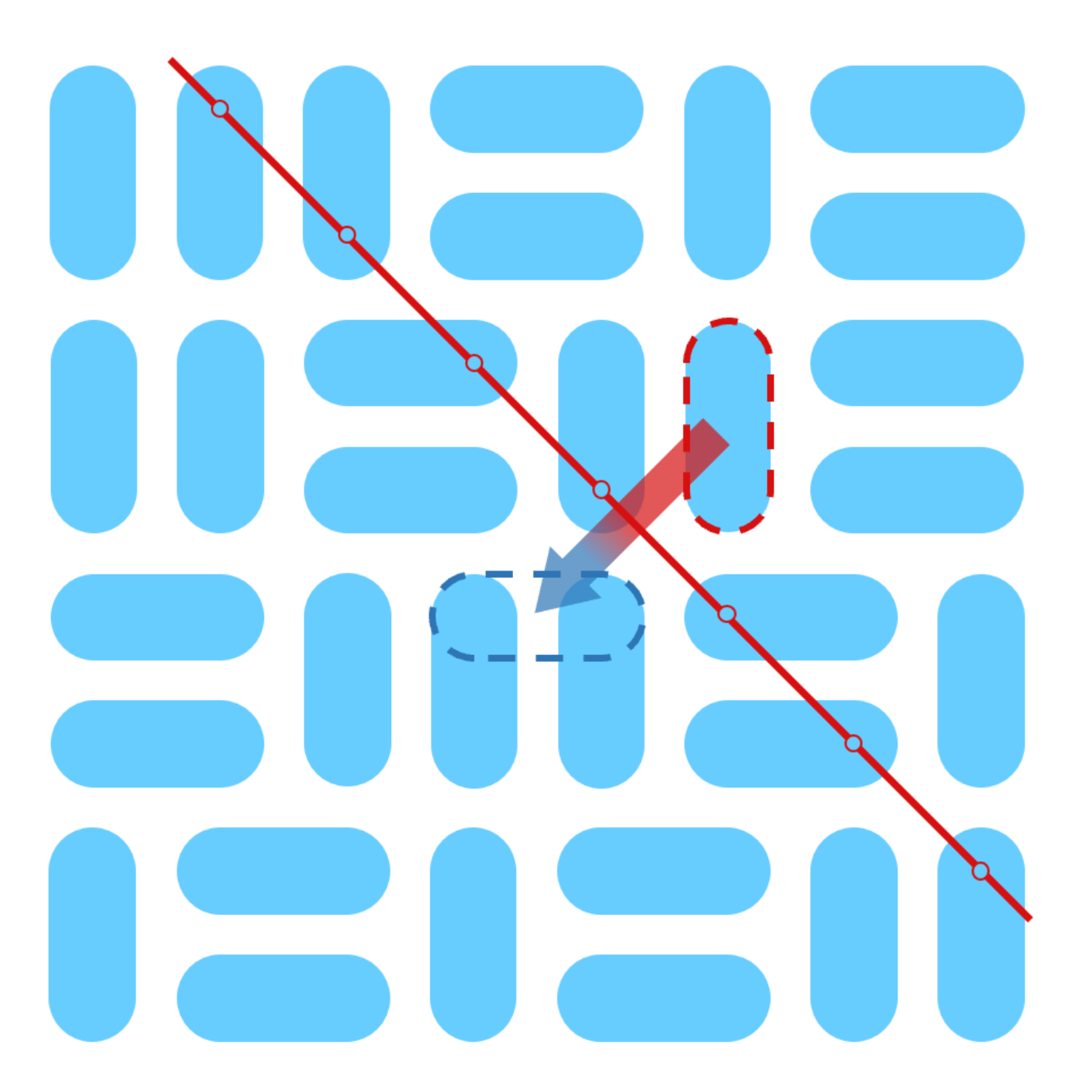}}
		\quad
		\subfigure[]{ \label{1f} \includegraphics[scale=0.25]{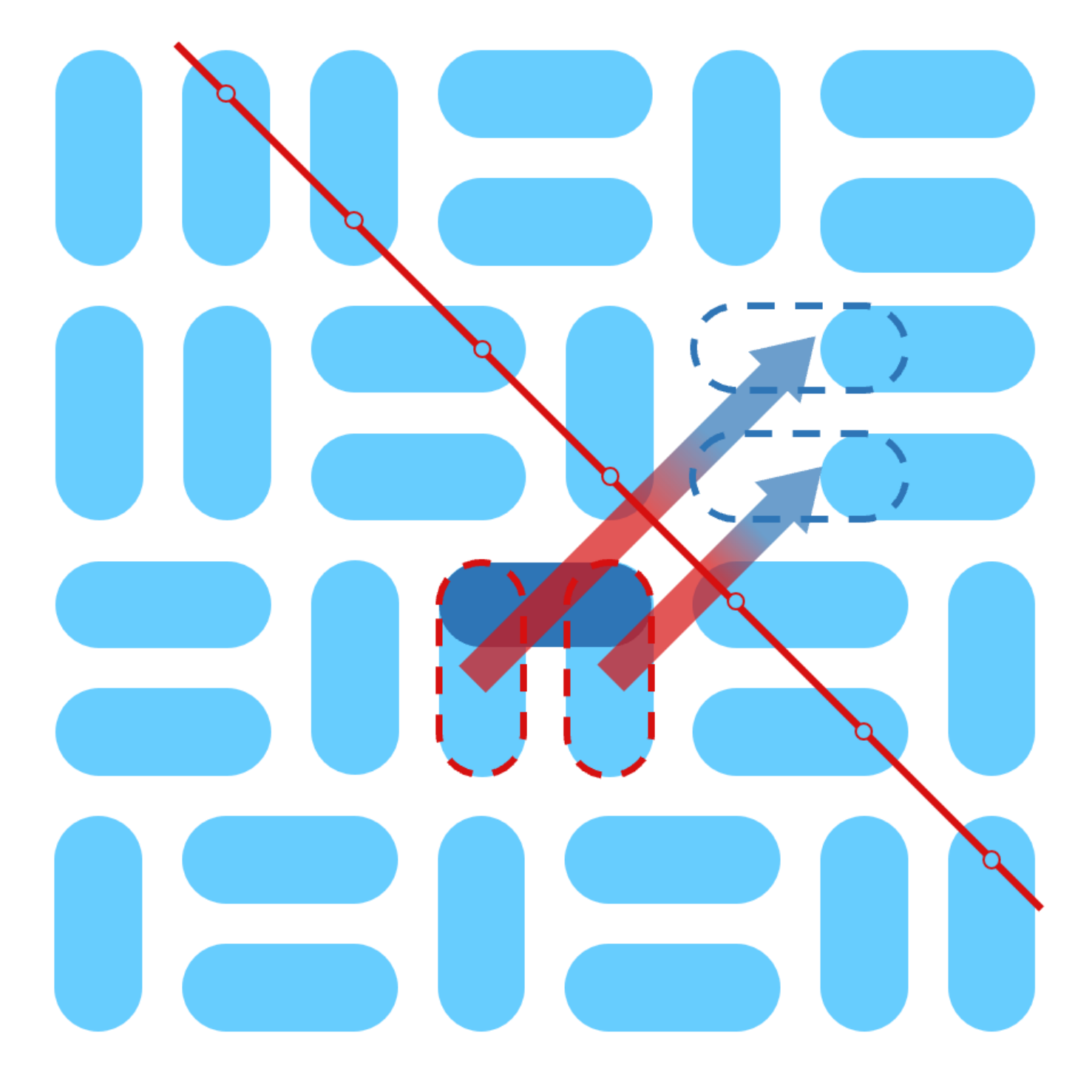}}
		\quad
		\subfigure[]{ \label{1g} \includegraphics[scale=0.25]{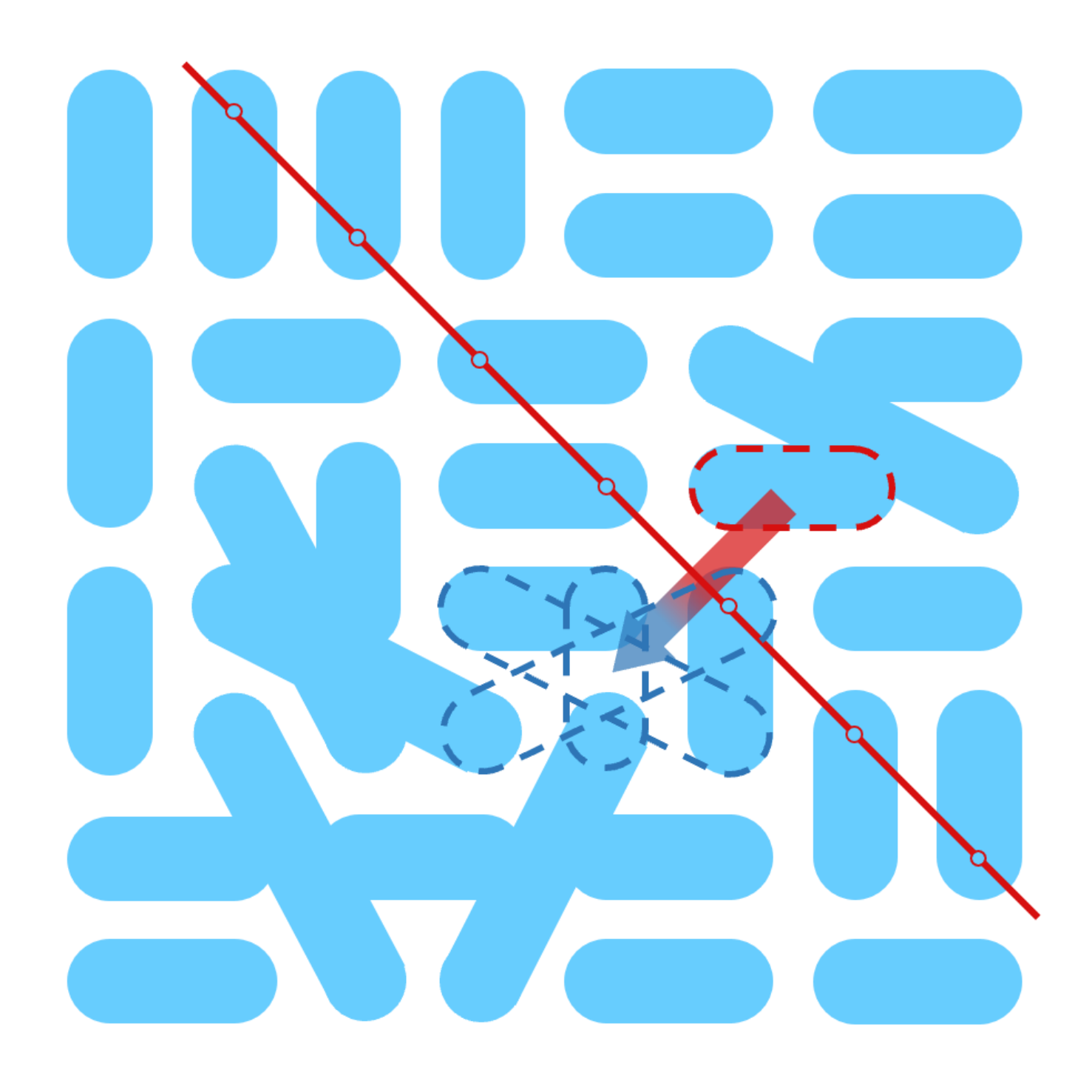}}
		\quad
		\subfigure[]{ \label{1h} \includegraphics[scale=0.25]{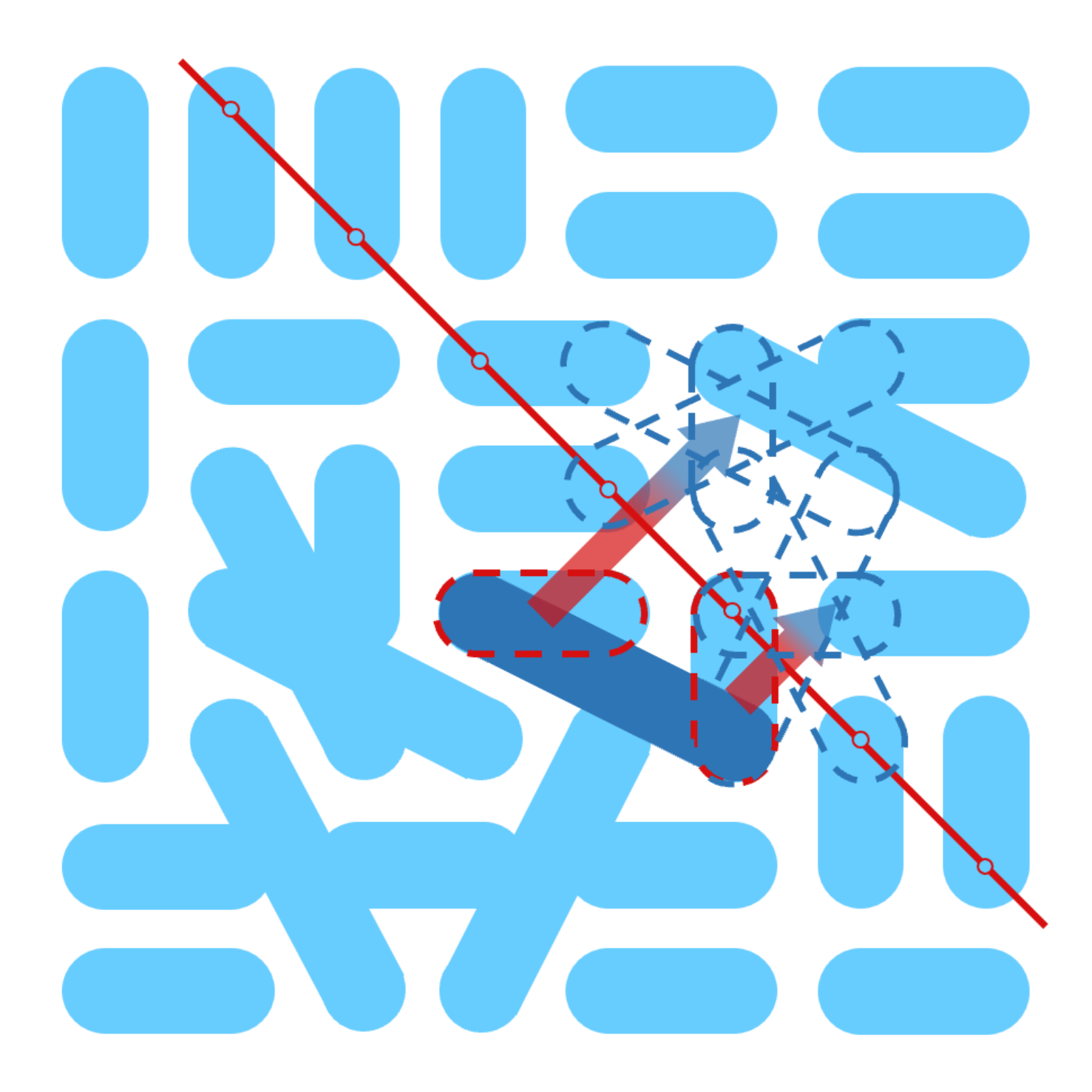}}
		\caption{(a)(b)(c)(d): schematic process of eDLA(blue point is the origin of the upcoming path and the red point is the directed point under energy criterion. Red dimer is the judging dimer and the blue dimer is the chosen dimer). (e)(f): schematic process of PLA on the close-packed dimer lattice. (g)(h): schematic process of improved PLA in SDM(In Fig.1.(e)(f)(g)(h), red line is the chosen axis of symmetry built on lattice points, red dashed lines cover the seed dimer and blue dashed lines cover mirror dimer).}		
	\end{figure}

\subsection{Pocket edge algorithm}
	PEA meliorated from the pocket Monte Carlo algorithm\cite{PhysRevB.67.064503} can also form long loops and traverse topological sections uniformly in close-packed dimers. The utilization of the symmetry axes and the introduction of the seed are two keys of our algorithm. In hard-core dimers, symmetry axes chosen in the algorithm will be introduced in Appendix A (see Fig.8).

	Here we broaden the using scope of PEA algorithm in SDM. Deconfinement of hard-core condition will increase the difficulty of graph theory and leverage us for understanding the generation and emergence of suchlike topological states or topological transition\cite{PhysRevB.54.12938,yan2021improved,yao2021breaking} such as Berezinskii-Kostelitz-Thouless (BKT) transition on the square lattice of hard-core dimers\cite{PhysRevLett.94.235702}.

	The formation process is in the following:

	1). Chose a symmetry axis randomly in the lattice. (four kinds of axes are figured in Fig.8(b) and here suppose it is the red line in Fig.1(e)(f)).  

	2). We choose a dimer as a seed dimer randomly (we wrap chosen seed dimers with red dashed lines in the Fig. 1(e)(f)).

	3). We reflect the seed dimer with the selected symmetry axis. The original seed will leave a hole in its original location and generate a mirror dimer in the symmetry location. If the mirror dimer overlaps other dimers, these overlapped dimers will be chosen as new seeds and the mirror dimer will stay there as the updated dimer (we wrap mirror dimers with blue dashed lines in Fig.1(e)(f). We fill new updated dimers as deep blue line and arrows means the reflection process).

	4). The seeds reflect and overlap repeatedly until the final seed are reflected to itself, thus we derive a new loop shown in Fig.8(g).

	As for SDM we take N1-N4 model as the example without loss of generality. N1-N4 model is the deconfinement of hard-core condition and second next-nearest dimers can be accepted in these cases (see in Fig.8(1)). The whole process is similar but we need to introduce an extra step to accomplish a multi-direction update. Only difference from above step aims at mirror dimers. Once the mirror dimer reflected by the seed dimer we add the following process:

	1). Once we derive a mirror dimer, we choose the center point of it and there are three dimers occupying the center point. In Fig.1(e)(f), we wrap mirror seed with blue dashed line.  

	2). We take these three dimers as our mirror dimers together, and randomly choose one of the three as the final mirror by possibility. Deep blue line marks the final mirror dimer in Fig.1(e).

	3). We fix the final mirror dimer as the new updated dimer and choose occupied dimers as new seed implementing above until a lock loop as the previous case.

	Our introduction of center point of mirror seed gives us chance to propagate horizontal and vertical dimers to other oblique dimers so that ergodicity can be achieved in SDM. The random choosing process around a center point guarantees detailed balance at the same time. Widening kinds of dimers around center points of mirror dimers can make us use the PEA in different dimers of deconfinement.

	Furthermore, we introduce multi-seeds simulation function to PEA to develop the efficiency. Due to the utilization of symmetry axes, PEA can form global loop based on the scale size of the model. It can also develop the simulation on large-scale lattice and traverse topological sections uniformly. Furthermore, we could use multi-seeds update in once simulation to achieve higher efficiency.

\section{NUMERIC RESULTS OF TESTING ALGORIHTMS}
	Our simulations are operated on $N = L \times L $ square lattices with periodic boundary conditions in Fig.8(a). From the previous study, the phenomenological Hamiltonian of QDM can be written as the following form\cite{PhysRevLett.61.2376,PhysRevB.54.12938}:
	\begin{equation}
		H =  \sum_{plaquettes}[-J(\ket{\mathop{\rule[0.1pt]{0.23cm}{0.05cm}}^{\rule[-0.1pt]{0.23cm}{0.05cm}}}\bra{\rule[-0.6pt]{0.05cm}{0.23cm}            \kern0.08cm\rule[-0.6pt]{0.05cm}{0.23cm} }+H.c.)+V(\ket{\mathop{\rule[0.1pt]{0.23cm}{0.05cm}}^{\rule[-0.1pt]{0.23cm}{0.05cm}}}\bra{\mathop{\rule[0.1pt]{0.23cm}{0.05cm}}^{\rule[-0.1pt]{0.23cm}{0.05cm}}}+\ket{\rule[-0.6pt]{0.05cm}{0.23cm}            \kern0.08cm\rule[-0.6pt]{0.05cm}{0.23cm}}\bra{\rule[-0.6pt]{0.05cm}{0.23cm}\kern0.08cm\rule[-0.6pt]{0.05cm}{0.23cm}}] 
	\end{equation}

	Where $J$ and $V$ are regarded as coupling constants, more detailed, we consider $J$ as interacting intensity and $V$ as potential intensity. $\rule[-0.6pt]{0.05cm}{0.23cm}            \kern0.08cm\rule[-0.6pt]{0.05cm}{0.23cm}$  and $\pmb{=}$ represent parallel dimers in a plaquette. For classical cases, we can let $J$ be zero. Thus, we can compute the total energy of each states by this Hamiltonian.

	Except this, we need an order parameter to represent the symmetry evolvement sensitively\cite{PhysRevB.54.12938,PhysRevLett.94.235702}. Based on their definition of order parameter, we define 2-dimensional vector order parameter $(\mu_x,\mu_y )$ on bipartite lattices including four-fold degeneracy of columnar ground states distinguishing from phases differences shown in Fig.2(a)(b)(c)(d).
	\begin{equation}
		\mu_x=\sum_{i,j}\epsilon^{ij}n_{ij}(-)
	\end{equation}
	\begin{equation}	
		\mu_y=\sum_{i,j}\epsilon^{ij}n_{ij}(\mid)  
	\end{equation}

	Where $\epsilon^{ij}=\begin{cases}
	1, \ if \ i \ is \ odd \ and \ j \ is \ even\\
	-1, \ if \ i \ is \ even \ and \ j \ is \ odd 
	\end{cases}$
	and $n_{ij}(-)=\begin{cases}
	1, \ if \ there \ is \ a \ dimer \ between \ vertex \ i \ and \ j\\
	0, \ if \ there \ is \ no \ dimer \ between \ vertex \ i \ and \ j
\end{cases}$
	It is same logic to $n_{ij}(\mid)$. For normalization, we compute the $\mu=(\mu_x^2+\mu_y^2)^{1/2} $as our absolute order parameters.

	For a square lattice with periodic boundary condition, every subspace could be a cut-off in Hilbert space and bases of each subspace can be orthogonal among different topological sections and traversing by global cross-over operation such as perfoliate loop like Kitaev toric code model\cite{KITAEV20032}. We can define winding number vector $(W_x, W_y)$, where $W_x=N_Y(A)-N_Y(B)$ and $W_Y=N_X(A)-N_X(B)$ shown in Fig.2(e)(f)).

	\begin{figure}[htbp]
		\centering
		\subfigure[]{ \label{2a} \includegraphics[scale=0.18]{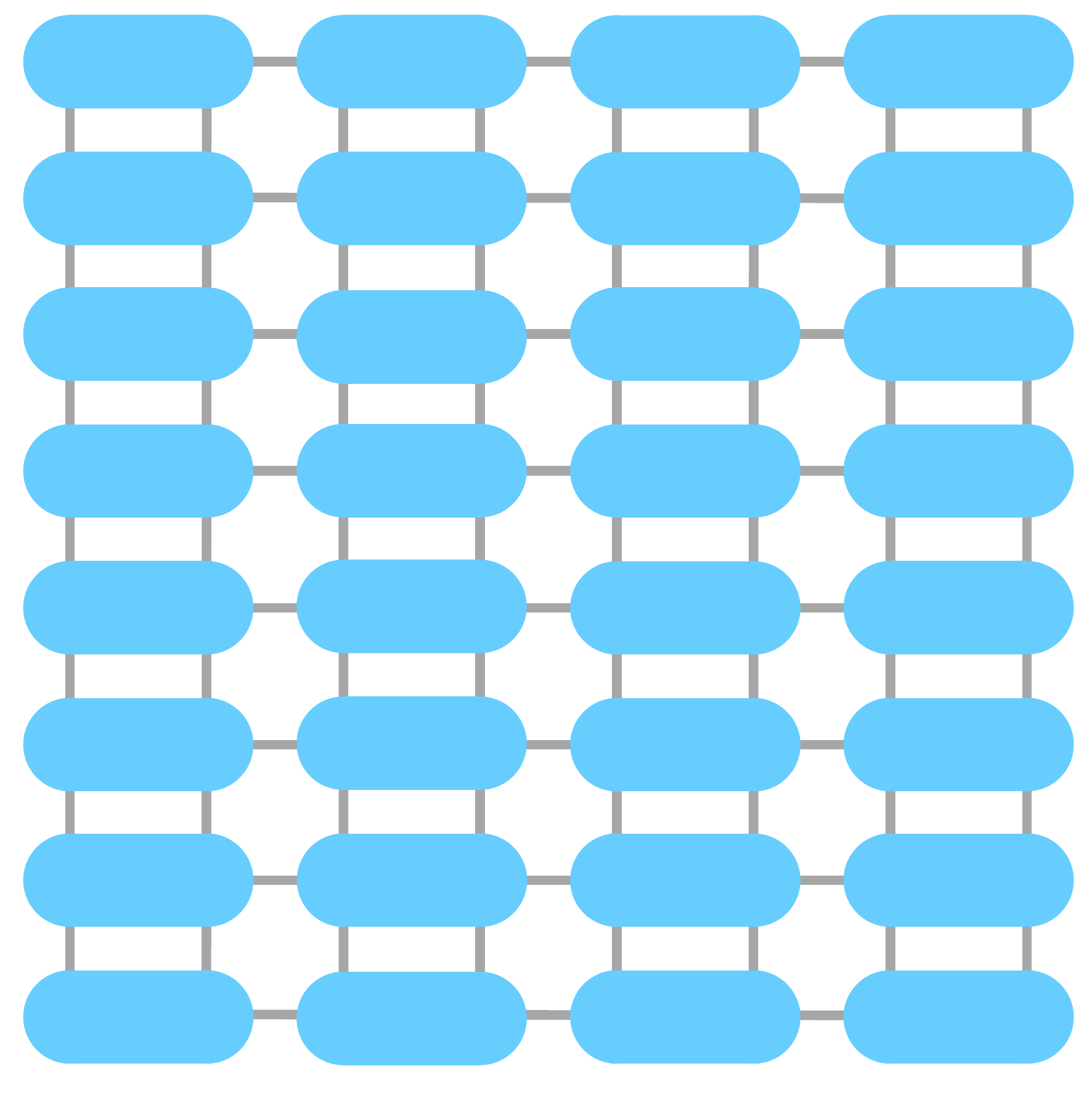}}
		\quad		
		\subfigure[]{ \label{2b} \includegraphics[scale=0.18]{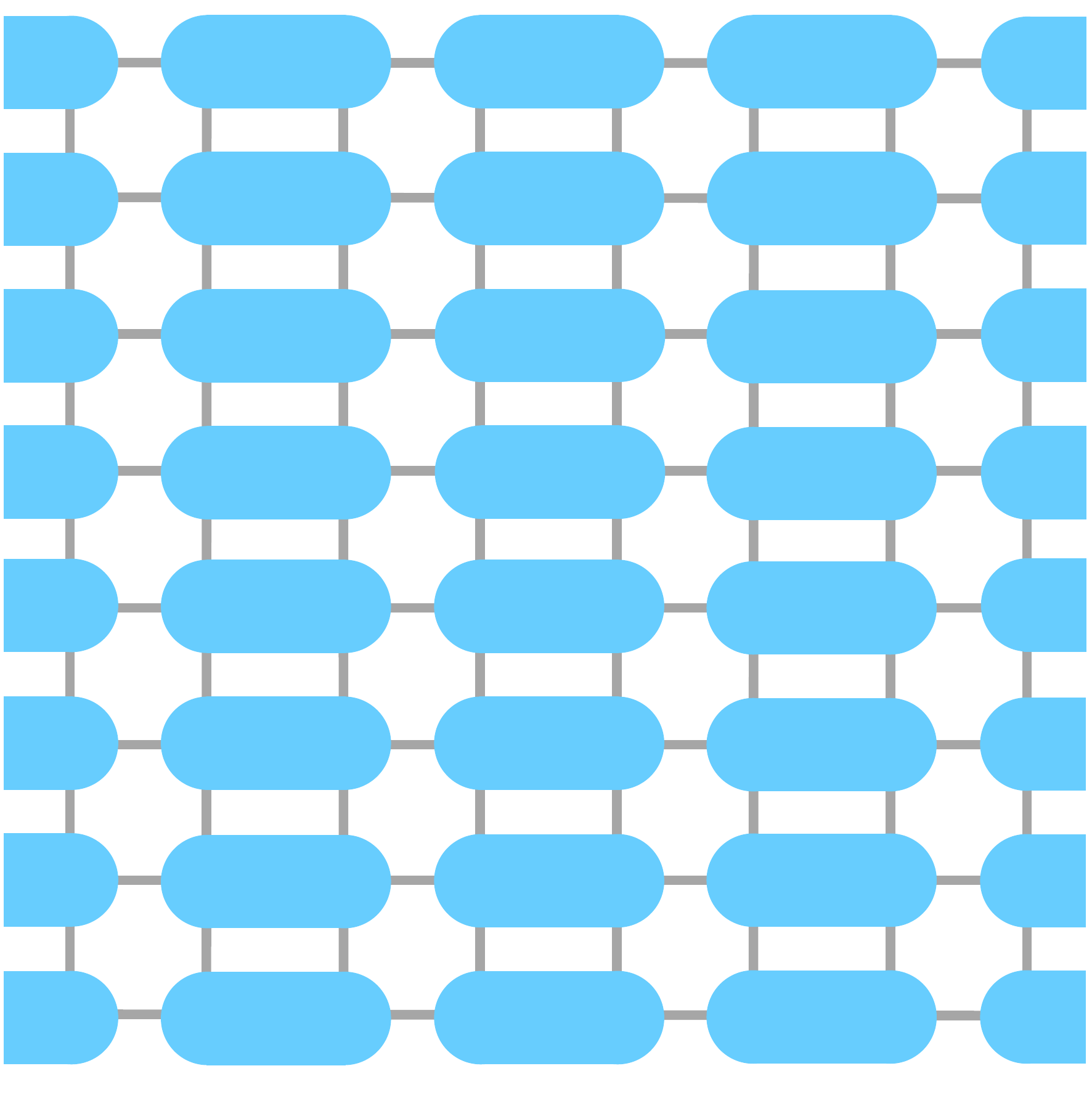}}
		\quad	
		\subfigure[]{ \label{2c} \includegraphics[scale=0.18]{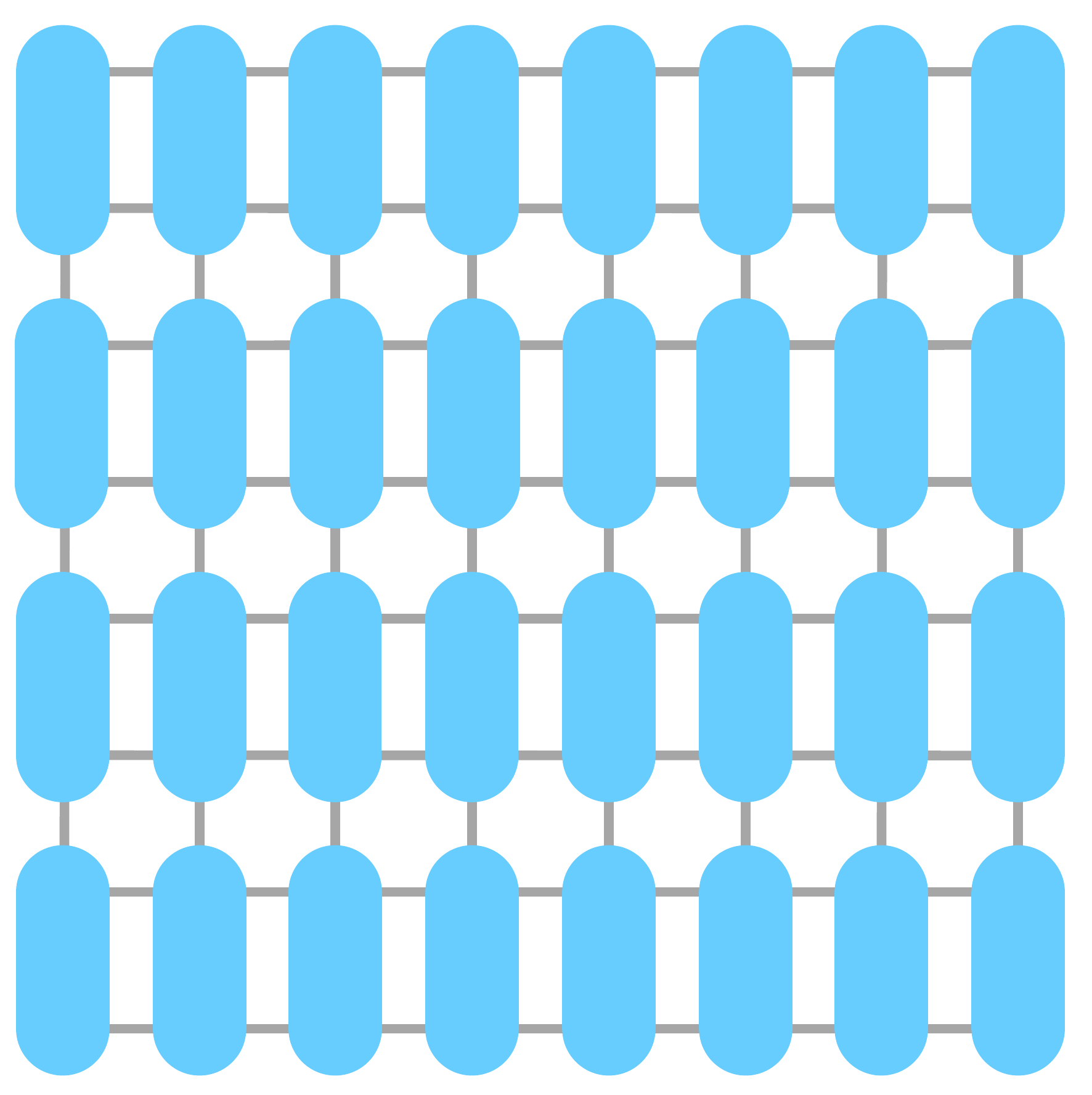}}
		\quad
		\subfigure[]{ \label{2d} \includegraphics[scale=0.18]{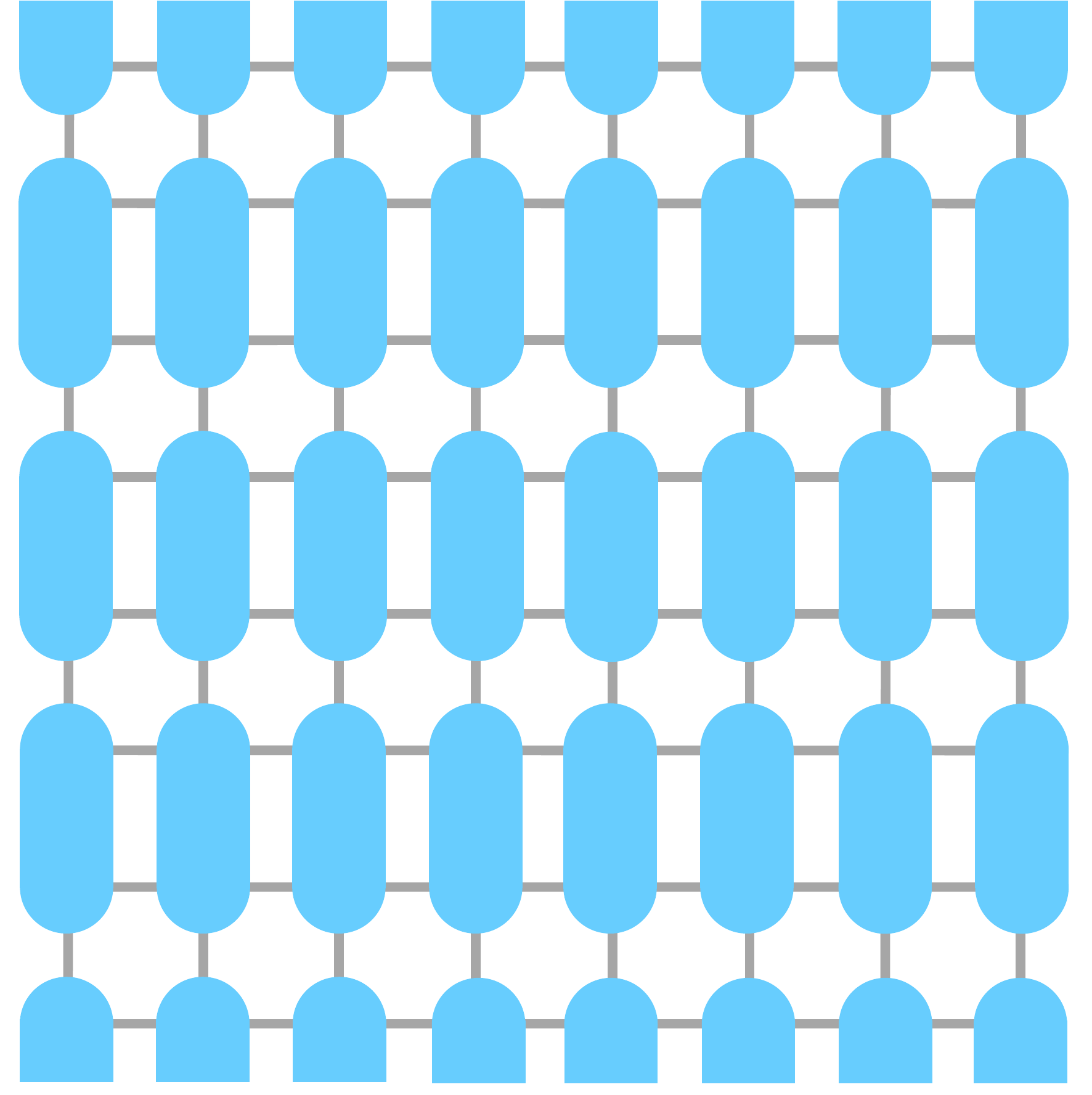}}
		\quad		
		\subfigure[]{ \label{2e} \includegraphics[scale=0.30]{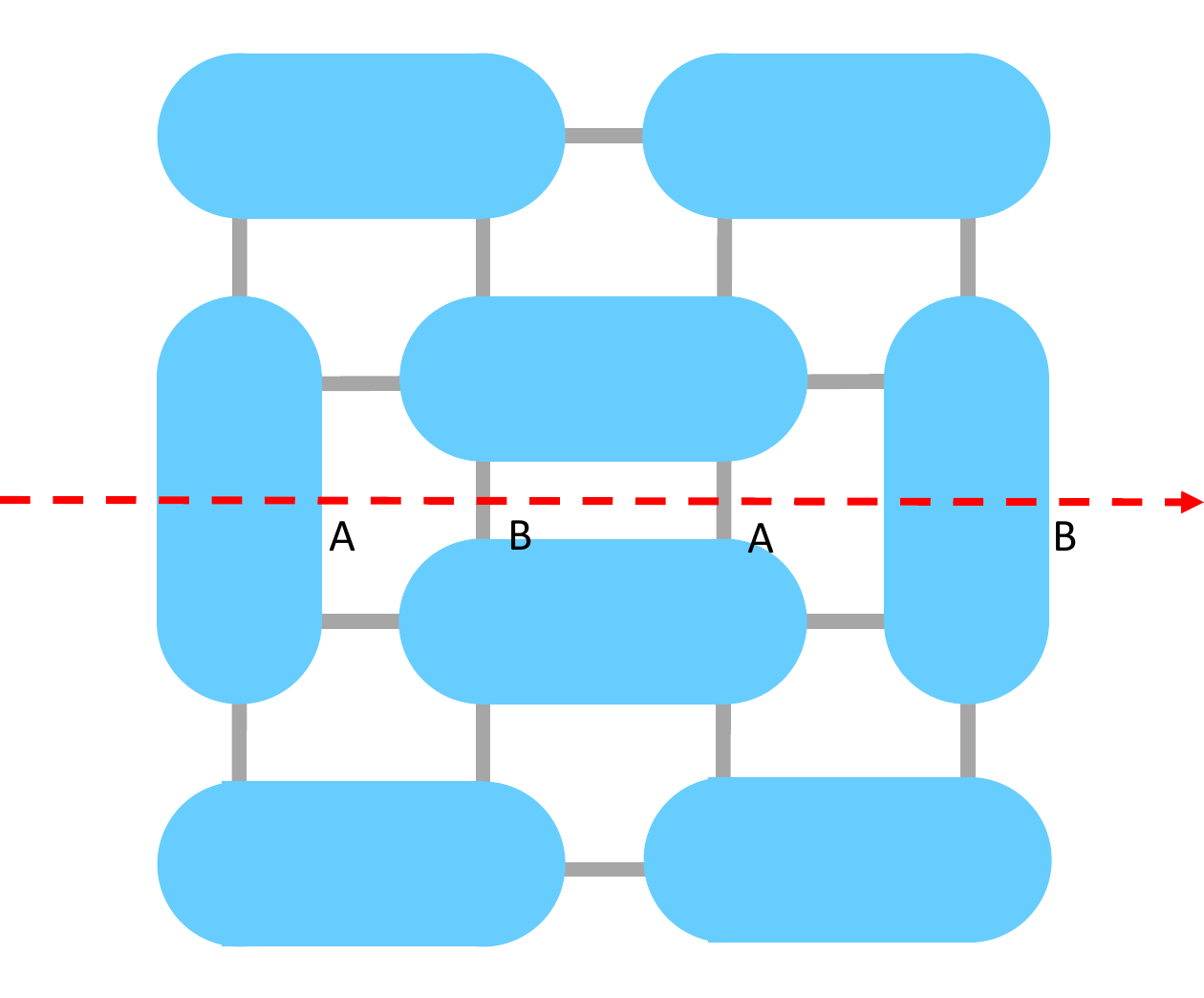}}
		\subfigure[]{ \label{2f} \includegraphics[scale=0.23]{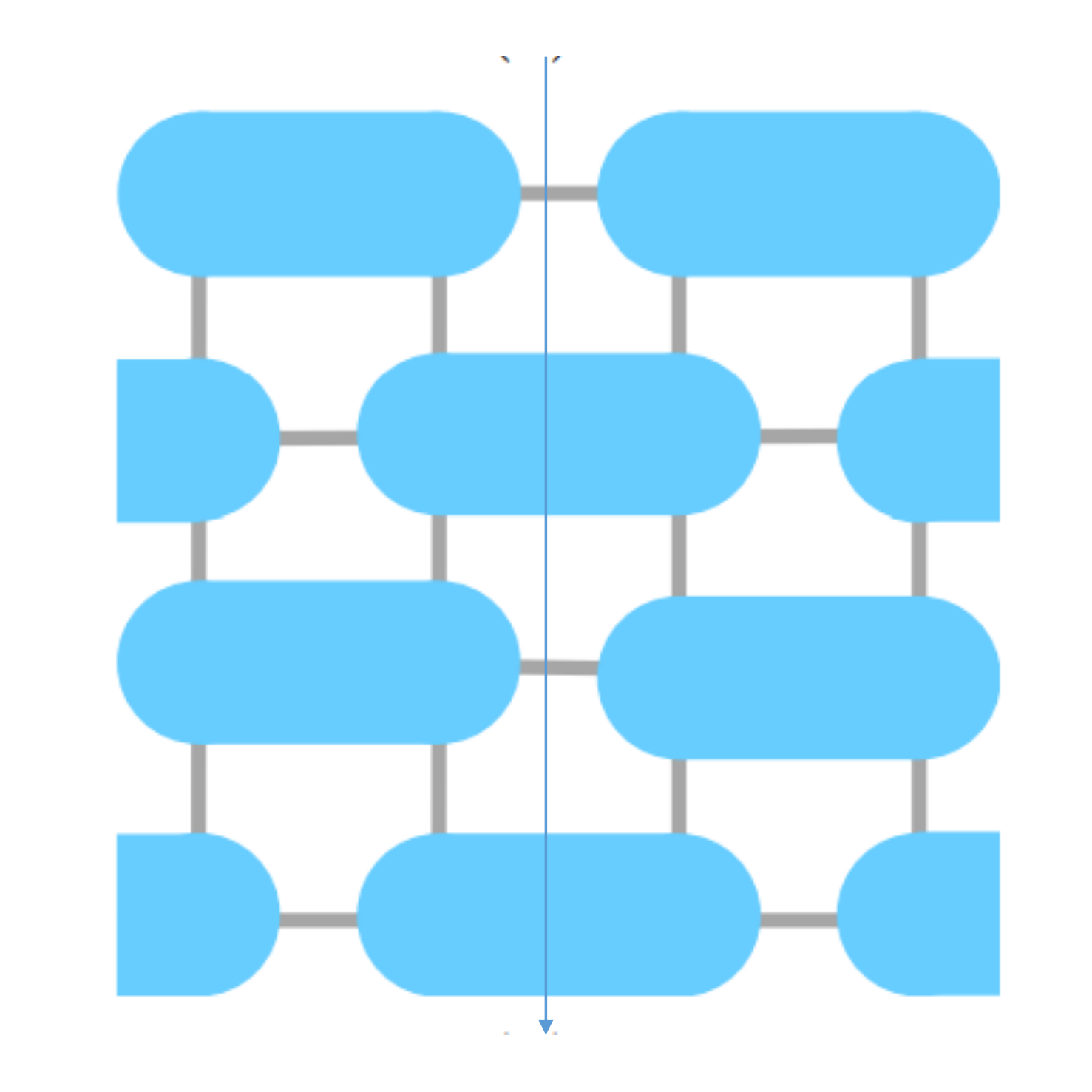}}
		\caption{(a)(b)(c)(d): four-fold degenerate columnar states in hard-core dimers on the square bipartite lattice. (e)(f): schematic computation of winding number $W_x$ and $W_y$. In Fig.2(e), we let a vertical line cross over from boundary to boundary and mark the edge as A, B, A, B one by one (we do not care the first edge is $A$ or $B$ and conclusions are the same) . $N_y(A)$ is the number of vertical dimers located on $A$ edges and $N_y(B)$ is the number of vertical dimers located on $B$ edges. Finally,$W_x=0$ in the figure. In Fig.2(f), we compute winding number $W_Y$ similarly and $W_Y= 2$ in this case.}		
	\end{figure}

	Firstly, we compare the convergent speed of these algorithms from an arbitrary degenerate ground states stable states on the lattice of $N=L\times L=64\times 64$ at different temperatures. In low temperature zone, system always breaks it symmetry according to Landau-Ginzburg phase transition theory such as Ising model\cite{RN18} or q-states Potts model\cite{PhysRevLett.62.361}. At low temperatures, degrees of freedom of the system will be frozen. All of algorithms are constrained and their abilities are difficult to be compared by order parameters. We can compute energy change and the acceptance ratio to analysis the behaviors of these algorithms at low temperatures (see Fig.3(a)(b)(c)).

	\begin{figure}[htbp]
		\centering
		\subfigure[]{ \label{3a} \includegraphics[scale=0.3]{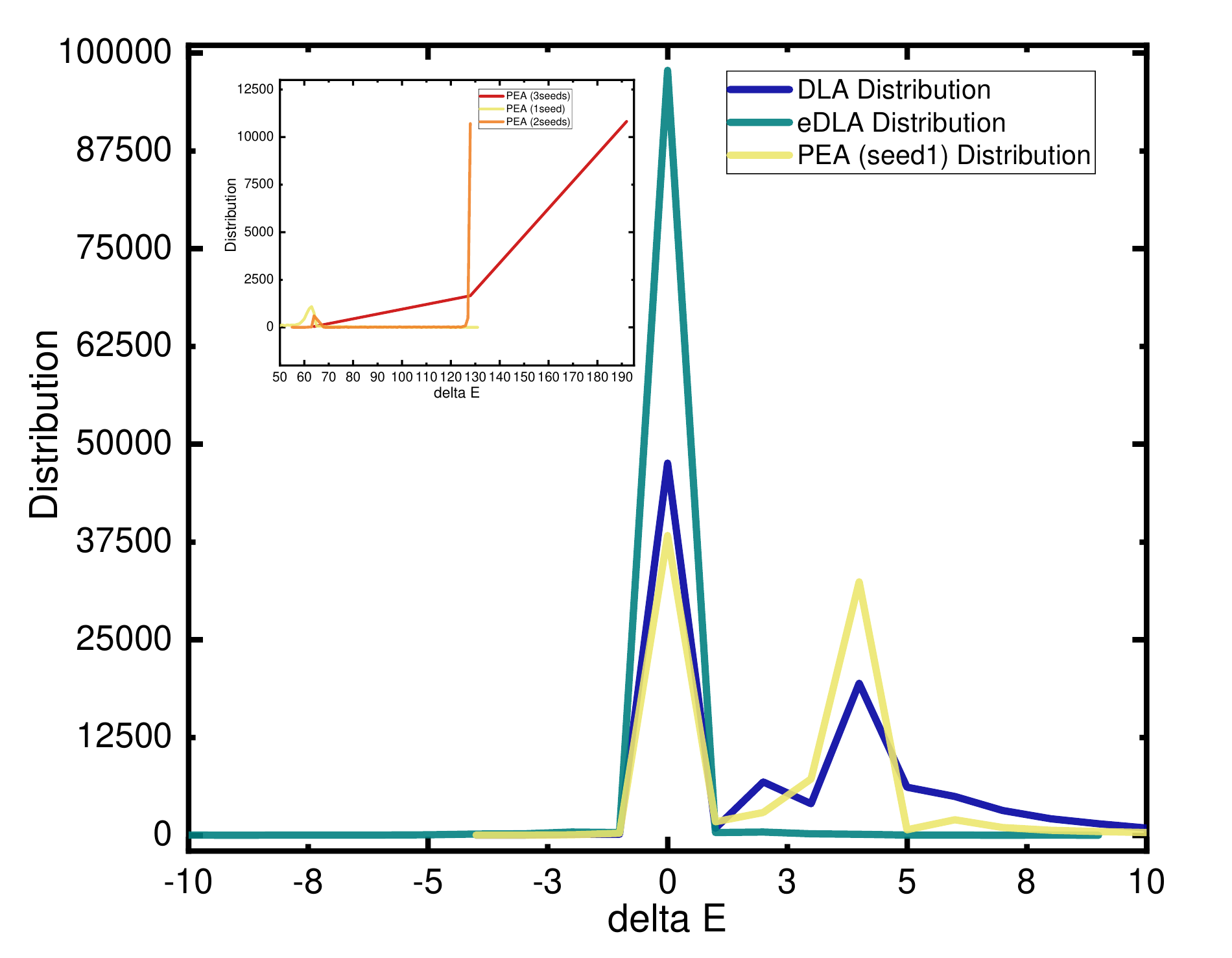}}
		\quad		
		\subfigure[]{ \label{3b} \includegraphics[scale=0.3]{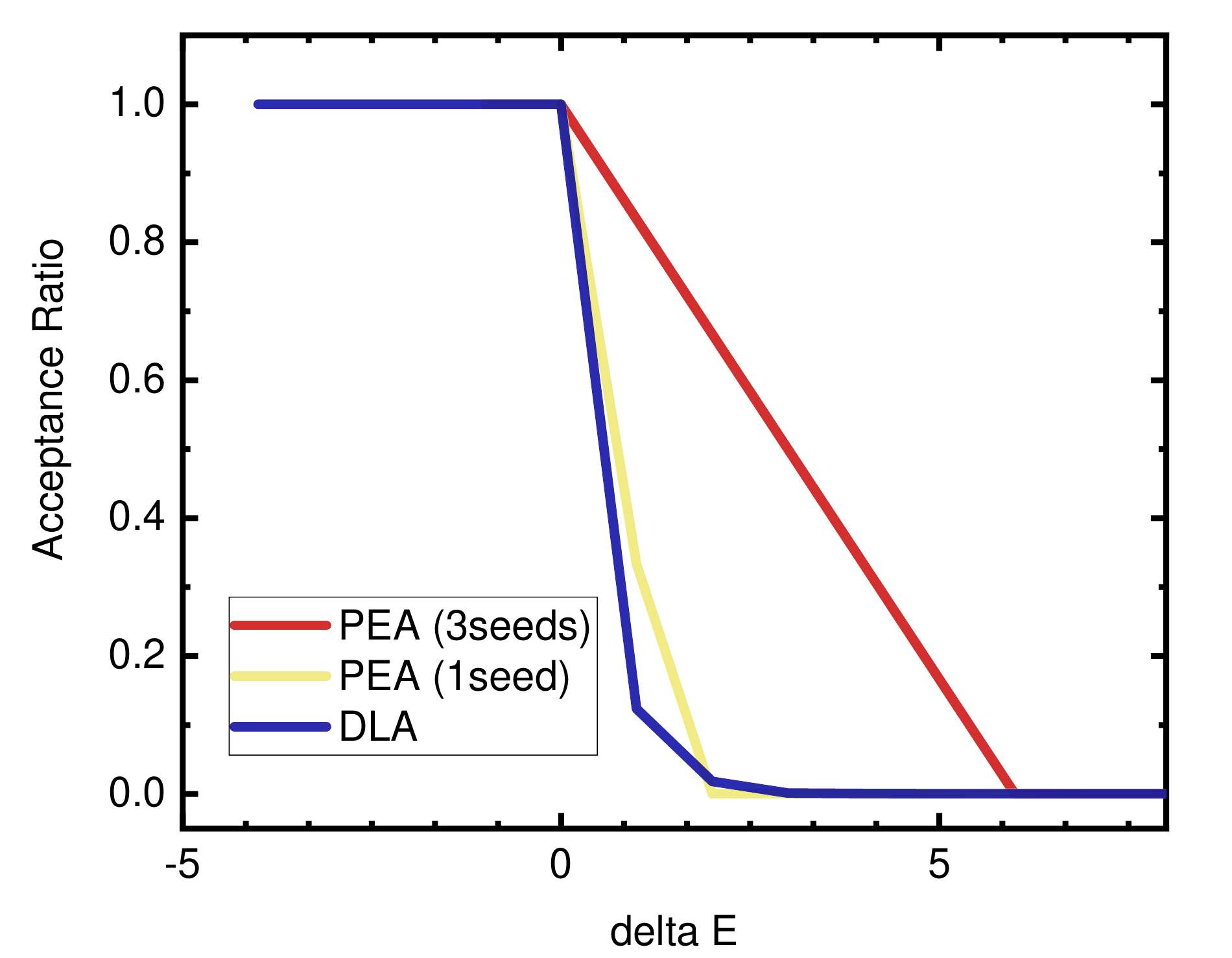}}
		\quad	
		\subfigure[]{ \label{3c} \includegraphics[scale=0.3]{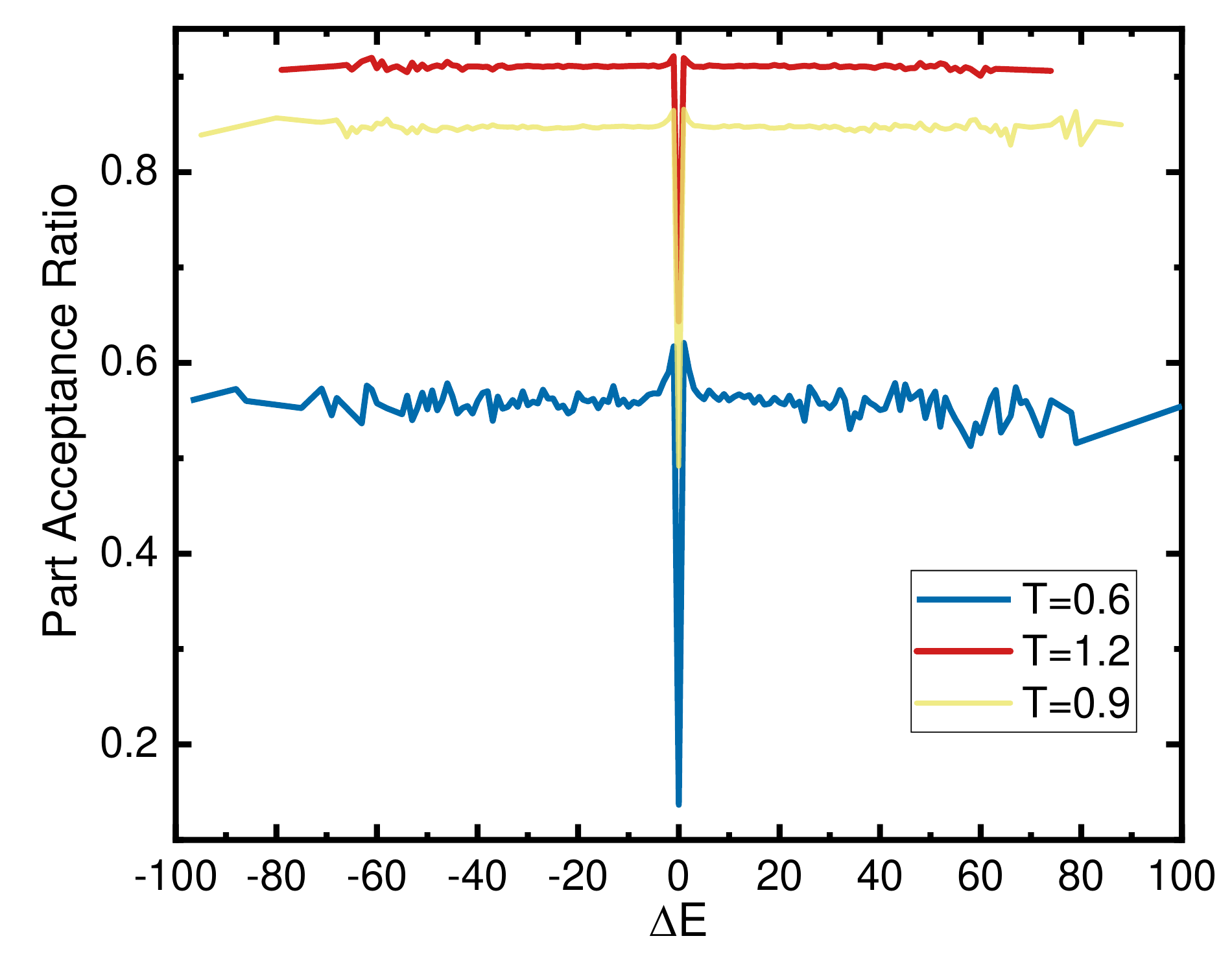}}
		\caption{(a):  Distribution of energy change of 100,00 times simulation of DLA, eDLA and PEA with 1 seed at $T=3.0$. (b): In Metropolis sampling process, acceptance ratio of DLA and PEA with 1 seed and PEA with 2 seeds. (c): part acceptance ratio of eDLA.}		
	\end{figure}

	Our study of three algorithms in low temperatures zone concentrates on the energy change $\Delta E$ distribution in 100,000 simulation times and compare acceptance ratio of their behaviors after relaxation. After 1,000,000 times simulation to relax, the system will enter into states possessing a characteristic energy scale and thermally fluctuate in this range. Talented algorithm at low temperatures will cause fluctuation as small as possible, so the system can derive thermodynamic quantities more accurately with small errors. Traditional DLA algorithm is a local algorithm, thus energy change $\Delta E$ will not be too large. For DLA, largest $|\Delta E|$ in 100,000 simulation times is 39. $\Delta E$ of eDLA simulation process gather around zero ($-10\le \Delta E\le 10$), that is, 98.88$\%$ loop can be formed to to causes energy change $-2\le \Delta E\le 2$ without strong fluctuations.  $\Delta E$ can hardly distribute more than +11 and less than -10. However, PEA with 1 seed can generate loops with $|\Delta E|$ more than 120 and if the energy change is more than +2 it can hardly be accepted (See in Fig.3(b)). $87.86\%$ loops formed by PEA with 1 seed gather in the interval of ($-4\le \Delta E\le 10$). More seeds mean larger energy change can be caused in Fig.3(a) subfigure and these loops are totally valid for the whole simulation process. Besides, DLA and PEA can generate secondary peak around $\Delta E\approx4$. Especially, PEA with more seeds can possibly generate tribble peak around $\Delta E\approx60$. Since every eDLA algorithm is not judged after a whole formed loop, we do not calculate acceptance ratio directly, every loops will be accepted as long as the complete loops are formed. Comparing loop acceptance ratio of DLA and PEA, on the right hand of $\Delta E=0$, lots of high-energy change loops can be accepted with higher possibilities and yellow and red acceptance ratio lines here describes the possibilities will rise up as the seeds increase for PEA (see Fig.3(b)). It means although PEA can form long loops with high-energy change, the generation of no-need loops of PEA is more than traditional DLA.  

	For eDLA, we define part acceptance ratio due to every loop formed by this method can be accepted. In the process of a loop in eDLA, energy paths can be passed or quitted. Those quitted paths will be a waste of computational power. We let efficient paths over total paths be our part acceptance ratio at three different temperatures ($T=0.6$, $T=1.2$,  $T=1.8$ and $T=3.0$). Higher temperature corresponding to higher thermally fluctuations means more paths can be accepted in a loop. At zero energy change point, there are valleys for part acceptance ratio because some of updates are starting from the origin point and back to the origin point. On the right and left hands of zero energy change point, there are two peaks of high efficient energy change points $\Delta E=+1$ and $\Delta E=-1$. These two peaks will evolve to none as temperatures rise up with stronger thermal flucturations.

	Except this, we need to focus more on the zone after phase transition. On square lattice with hard-core dimers, the thermodynamic transition is a BKT transition. It means the whole transition process is tempered that yields the change of order parameter is not obvious approaching to the transition temperature $T_c$ and in a short temperature zone after $T_c$. According to our other research, $T_c\approx 0.6$ can be derived\cite{yao2021breaking}.  Thus, we describe three curves at three characteristic temperatures ($T=1.2, T=3.0$ and $T=inf$) away from $T_c$ gradually. 

	\begin{figure}[htbp]
		\centering
		\subfigure[]{ \label{4a} \includegraphics[scale=0.3]{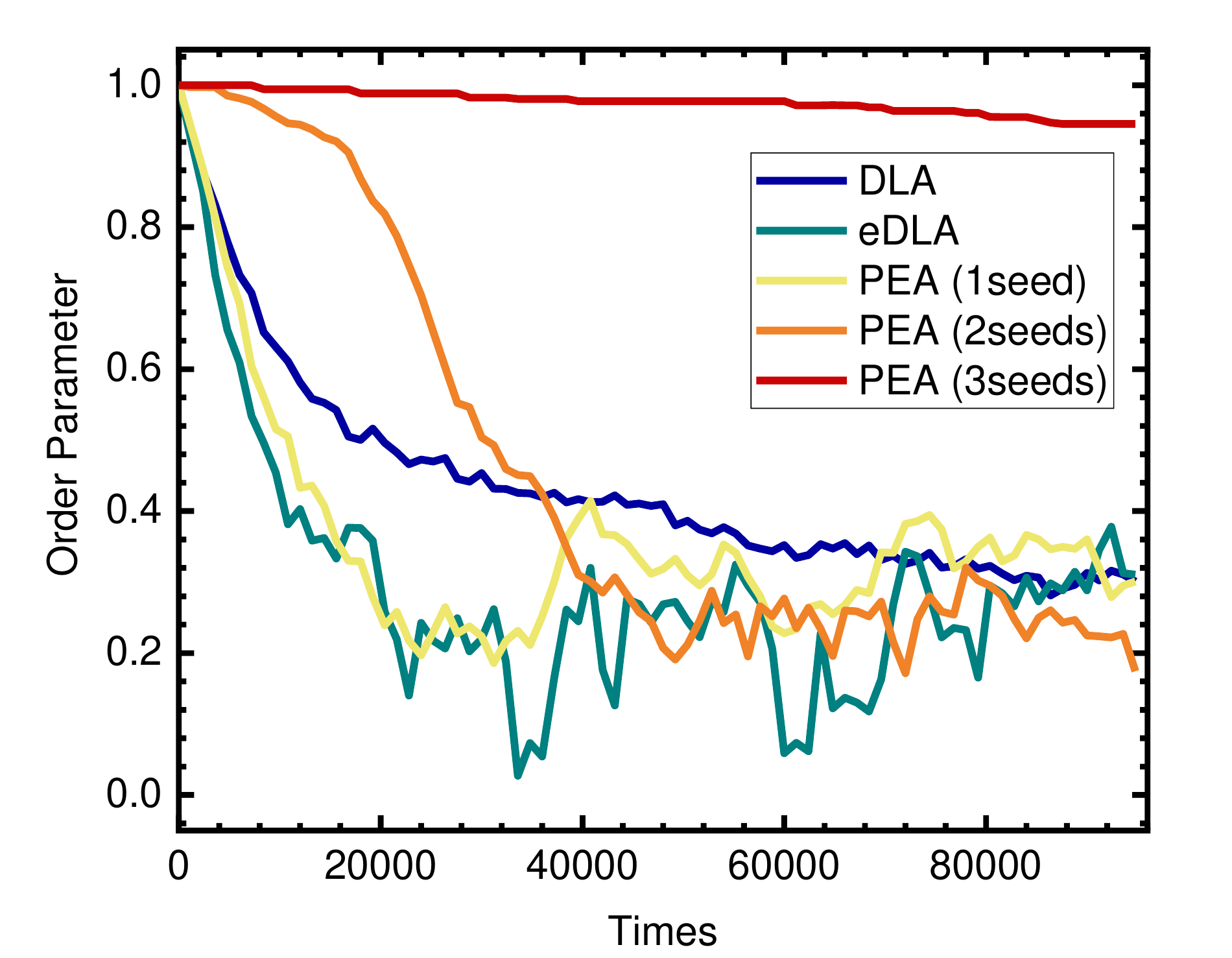}}
		\quad		
		\subfigure[]{ \label{4b} \includegraphics[scale=0.3]{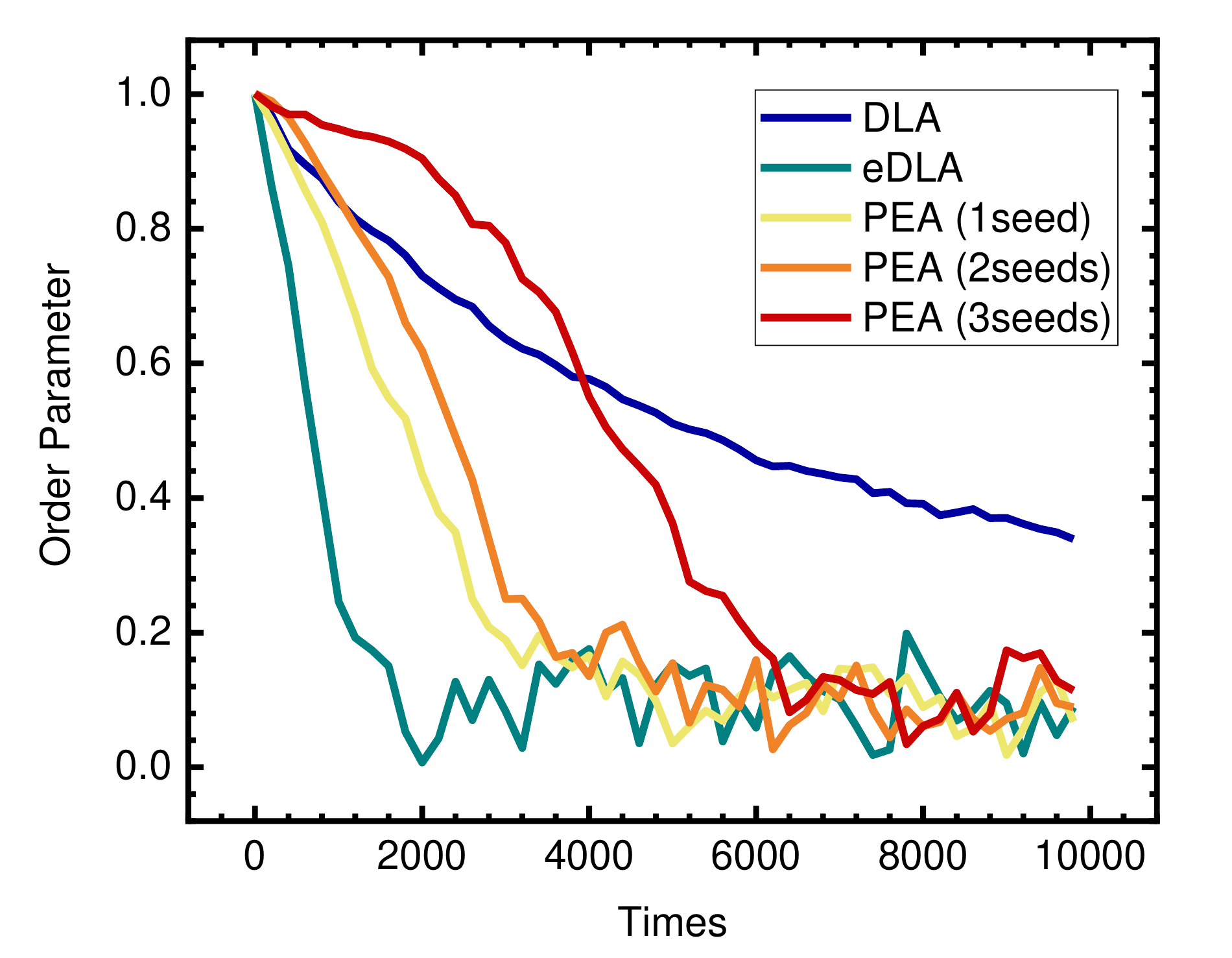}}
		\quad	
		\subfigure[]{ \label{4c} \includegraphics[scale=0.3]{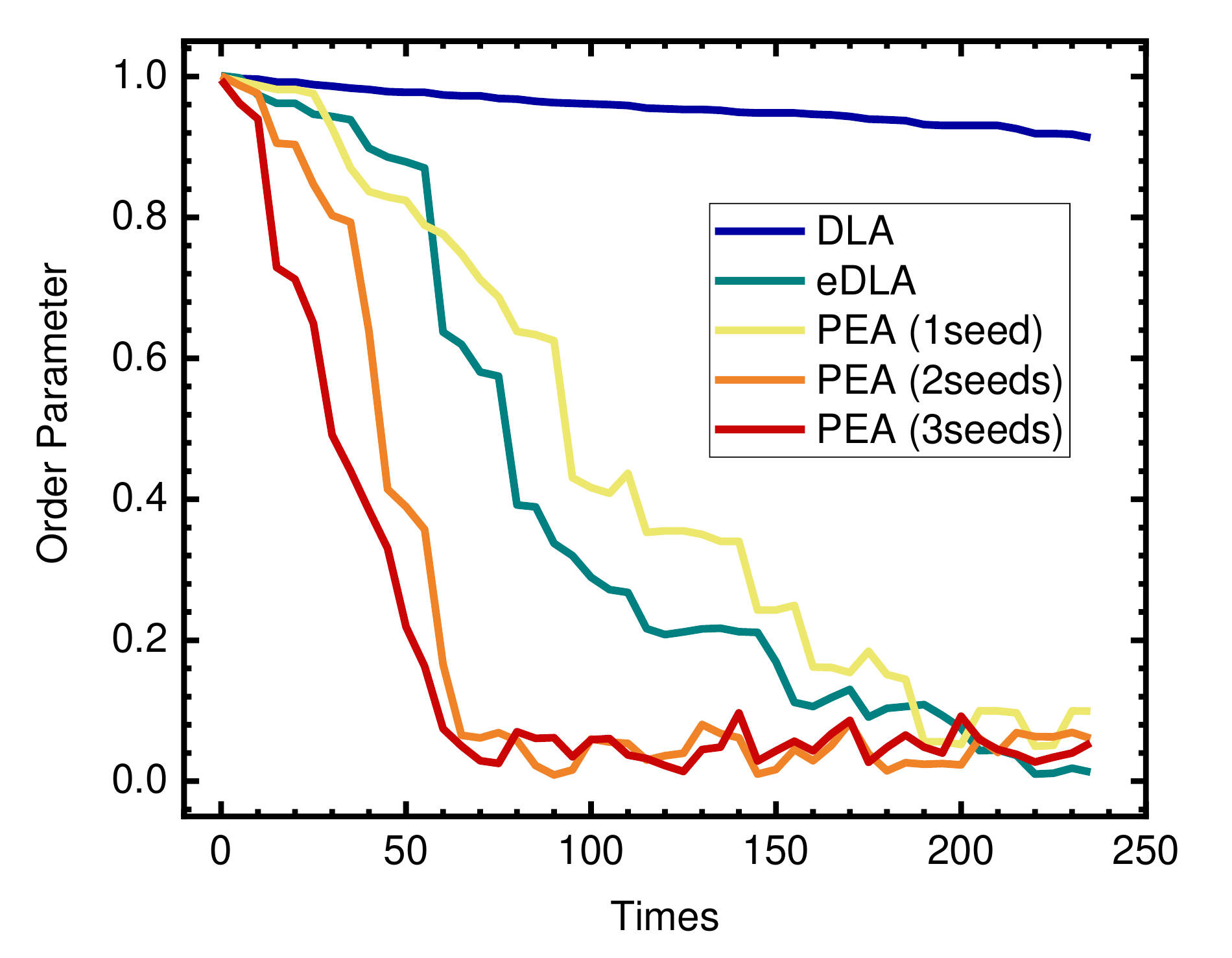}}
		\caption{(a): At $T=1.2$, relaxation processes by five methods (DLA, eDLA, PEA with 1 seed, PEA with 2 seeds and PEA with 3 seeds). (b): $T=3.0$. (c): $T=infinity$.}		
	\end{figure}

	In our simulation tests, in high temperatures zone, the convergent speed of eDLA is fastest among five methods shown in Fig.4(a)(b). From two figures, we can easily see convergent speed of eDLA is faster than others obviously. Accurately, we compute convergent time $\tau_{con}$. We fit convergent curve with exponential function $\alpha exp(-t/\tau_{con})+\beta $. Here we use more seeds to simulate at high temperatures. The convergent time $\tau_{con}$ of DLA, eDLA, PEA with 1 seed, PEA with 2 seeds and PEA with 3 seeds can be fitted sequentially as $1/\tau_{con}^{DLA}  =0.002354$, $1/\tau_{con}^{eDLA}  =0.725639455$,$1/\tau_{con}^{PEA1}=0.005423652$, $1/\tau_{con}^{PEA2}=0.004233507$, $1/\tau_{con}^{PEA2}=0.00144092$. Convergent time of eDLA is shorter than others and PEA more seeds will achieve little. Since the system enters into a heat pool with infinite energy at infinite temperature, longest loops generating by PEA with seeds as many as possible impulse every energy change can be accepted and the whole system lives in state fluctuating over and over again. At this time, eDLA will lose its progressing criterion just like travelers without maps. But in finite-temperature simulation, it will not become our problem. Therefore, we study the temperature dependence of convergent time $\tau_{con}$. Below $T=0.8$ for PEA with 2 seeds and below $T=1.2$ for PEA with 3 seeds, convergent processes are too difficult to occur due to the long loops do not be accepted. For classical hard-core dimers, $T=3.0$, the system will enter into non-order states and the order parameter $<\mu>\approx 0.0095\sim0$.

	We can calculate the temperature dependence of convergent decaying time and the scale dependence of loop length to consider the properties of locality and globality. Seeing in Fig.5(a), $1/\tau_{con}$ of eDLA is higher than others markedly. PEA with more seeds will ascend intensively as temperature rises and only temperature is higher enough that can transcend eDLA. Traditional DLA as a local algorithm do not appear apparent change and just more loops can be accepted. In our figures of scale dependence of loop length, we consider total loop length and efficient loop length simultaneously. Total loop length is to count average of loop length of all of loops and efficient loop length is to count only accepted loops. It remarks traditional DLA is a local algorithm and it will not increase as the scale increase. But eDLA will be definitely sharpen up as a global algorithm and the loops length of eDLA will grow as scales grow. Although total loops of PEA can grow as scales grow, too many energy-expensive loops will be formed. Thus, PEA as an intermediate non-local algorithm will not display too many advantages.

	\begin{figure}[htbp]
		\centering
		\subfigure[]{ \label{5a} \includegraphics[scale=0.3]{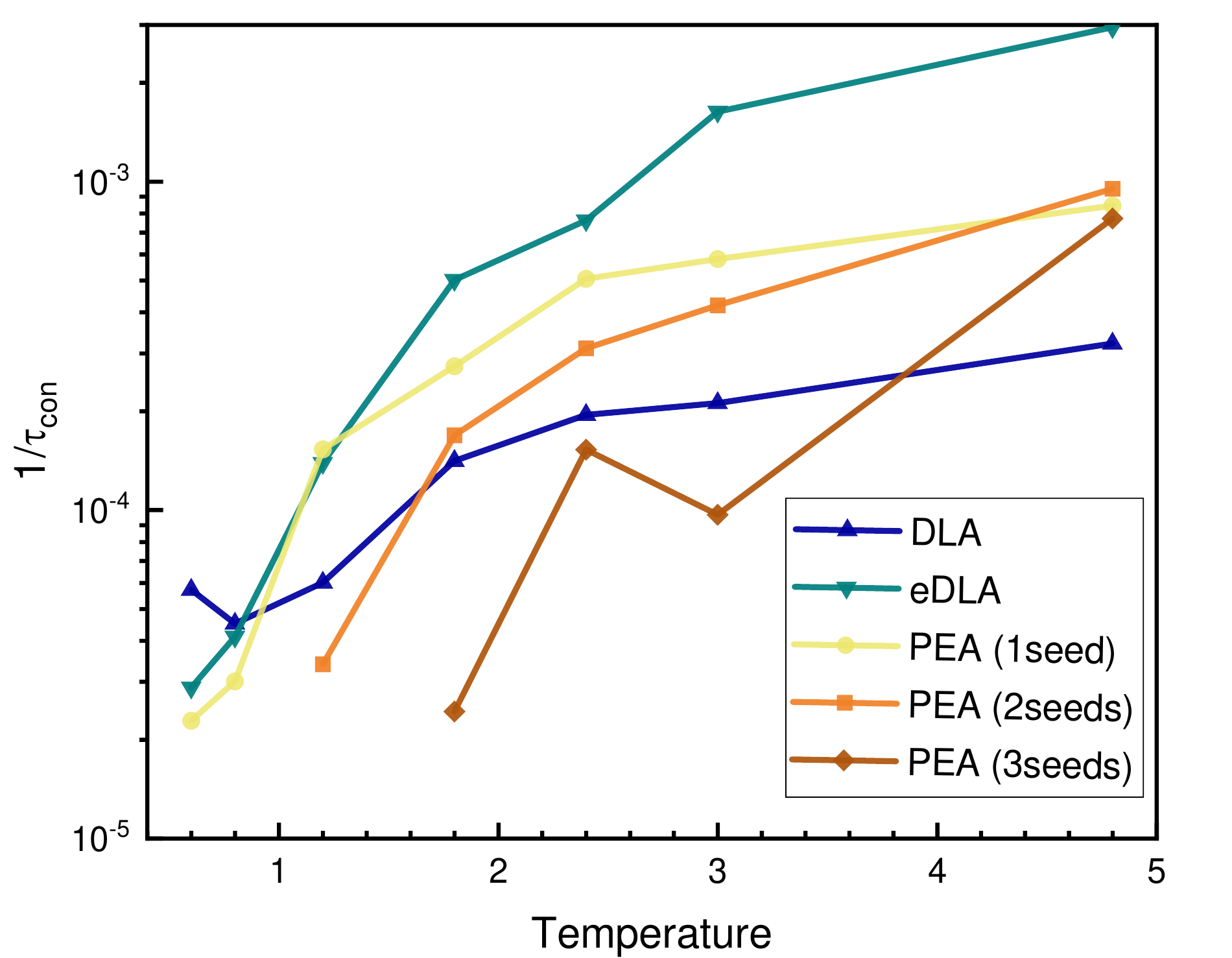}}
		\quad		
		\subfigure[]{ \label{5b} \includegraphics[scale=0.3]{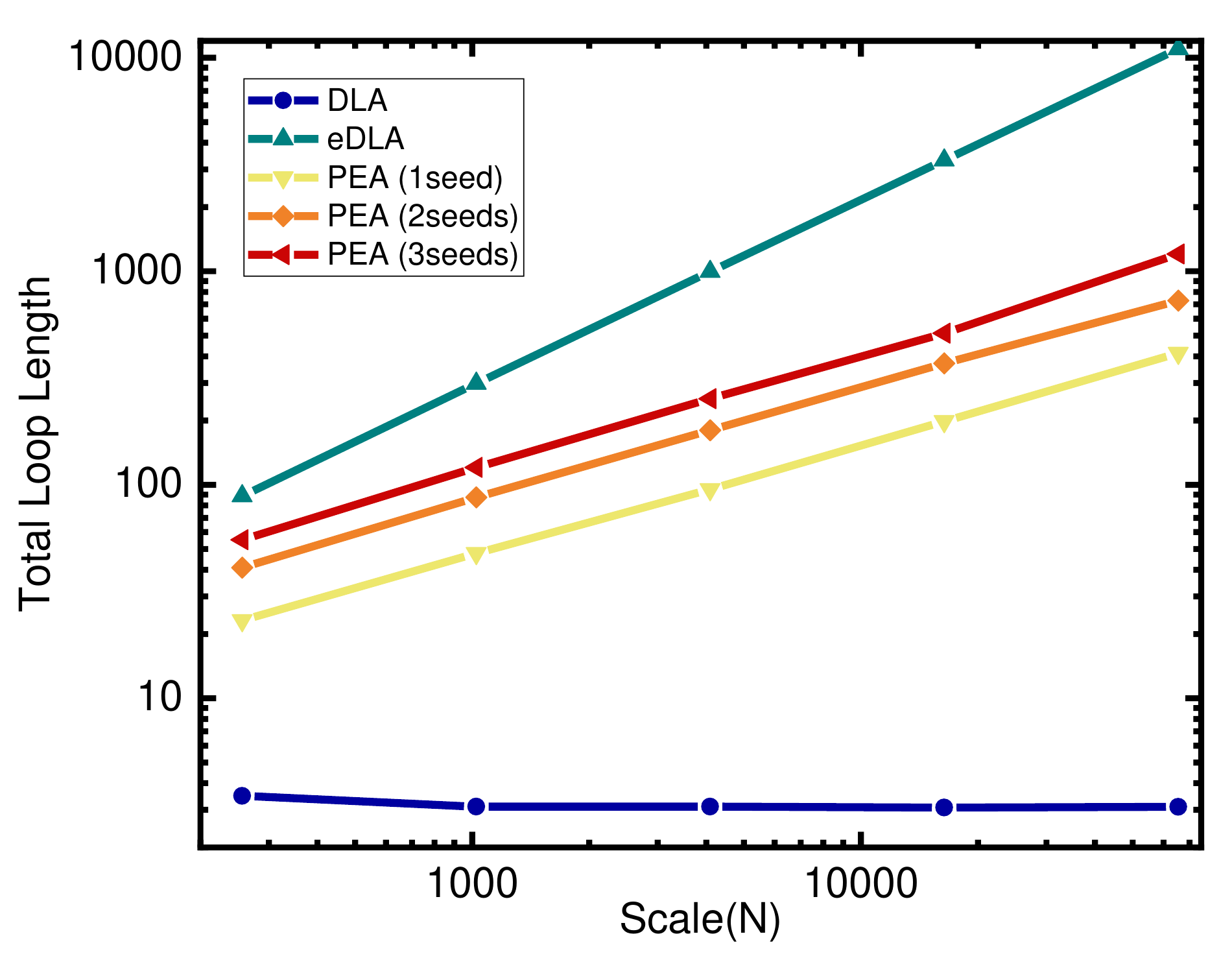}}
		\quad		
		\subfigure[]{ \label{5c} \includegraphics[scale=0.3]{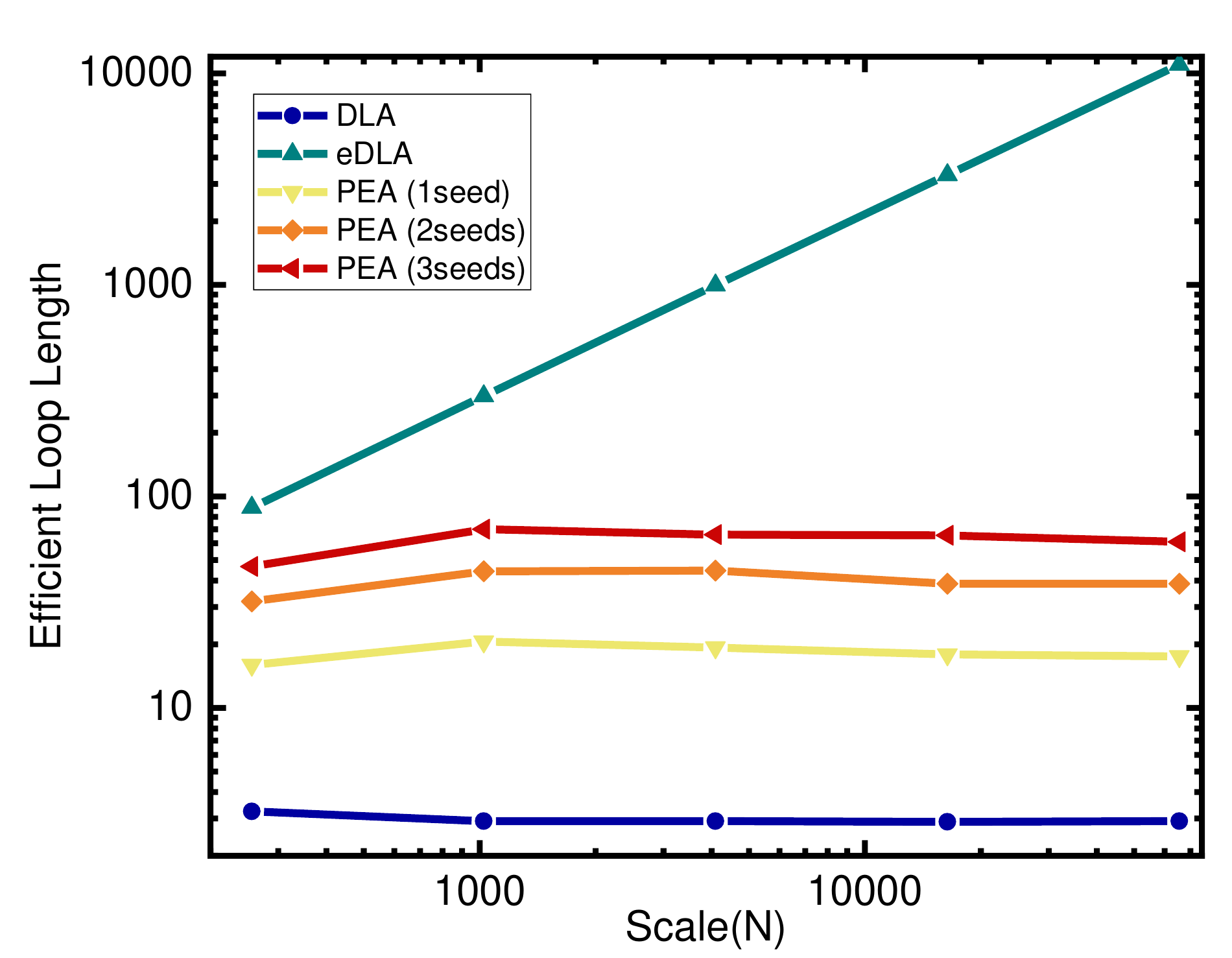}}
		\caption{(a): temperature dependence of convergent time by five methods (DLA, eDLA, PEA with 1 seeds, PEA with 2 seeds, PEA with 3 seeds). (b) scale dependence of loop length by five methods (DLA, eDLA, PEA with 1 seeds, PEA with 2 seeds, PEA with 3 seeds). (c): scale dependence of efficient loop length by five methods (DLA, eDLA, PEA with 1 seeds, PEA with 2 seeds, PEA with 3 seeds).}		
	\end{figure}

	\begin{figure}[htbp]
		\centering
		\subfigure[]{ \label{6a} \includegraphics[scale=0.3]{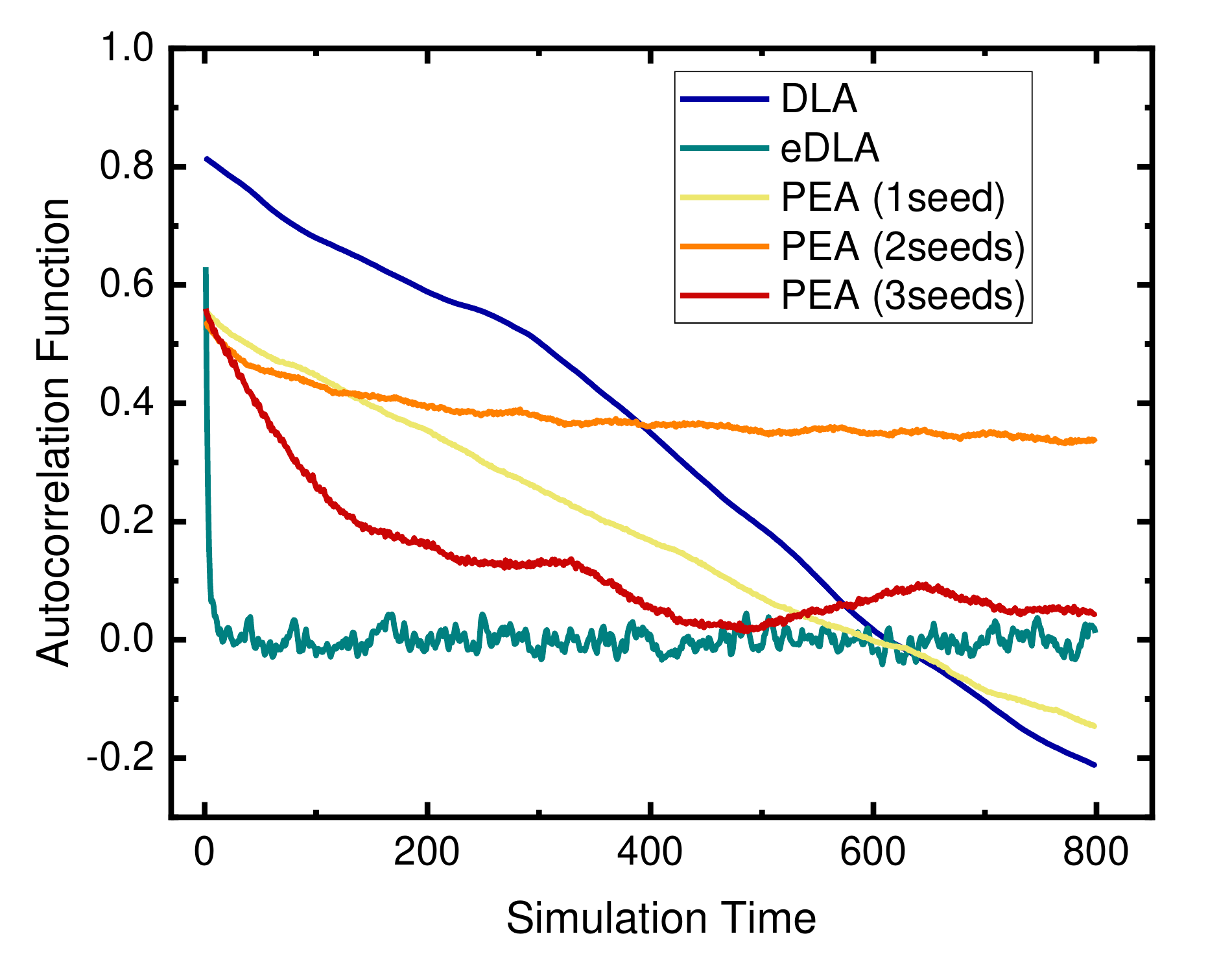}}
		\quad		
		\subfigure[]{ \label{6b} \includegraphics[scale=0.3]{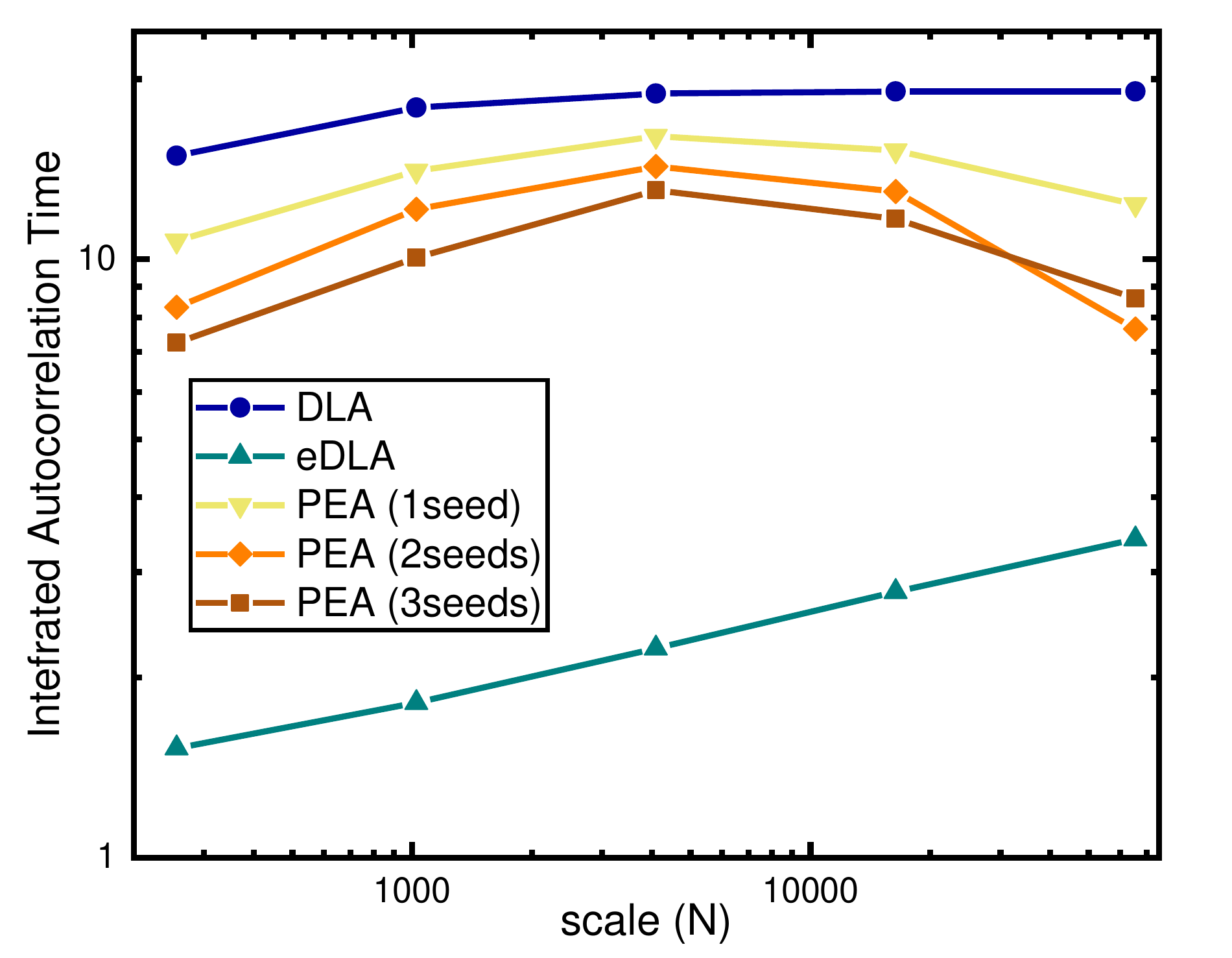}}
		\caption{(a): After relaxation, autocorrelation function of system ($L=64$) by methods of DLA, eDLA, PEA with 1 seed, PEA with 2 seeds and PEA with 3 seeds. (b): scale dependence of integrated autocorrelation time by methods of DLA, eDLA, PEA with 1 seed, PEA with 2 seeds and PEA 3 seeds.}		
	\end{figure}

	Furthermore, we study autocorrelation properties of three algorithms. We define autocorrelation function describes decorrelated measurements length\cite{2008LNP...739.....F} $C(t)=\frac{{\textstyle \sum_{i}^{N}O_{i}O_{i+t}}}{<O^2>}\times\frac{1}{N-t}$ where $O_i= \mu_i-<\mu>$. Clearly, due to a data must be autocorrelated with itself, $C(0)=1$. Calculating from step 1, for eDLA, its convergence tends to zero rapidest with shortest autocorrelated steps. PEA with different seeds occupy the meddle ground of local and global algorithms (see Fig.6(a)). In next step, we consider the integrated autocorrelated time $\tau_{int}=\frac{1}{2}+{\textstyle \sum_{i}}C(t)$ and measure the lattice scale dependence of autocorrelations. Traditional DLA shows strong autocorrelation no matter how large or small of the lattices. Because of large loops, PEA will be strong correlated when lattices are small and decrease its correlations when compute the lattices large enough. Improved eDLA with weak autocorrelations and they slightly increases as scale grows (see Fig.6(b)). 

	\begin{figure}[htbp]
		\centering
		\subfigure[]{ \label{7a} \includegraphics[scale=0.3]{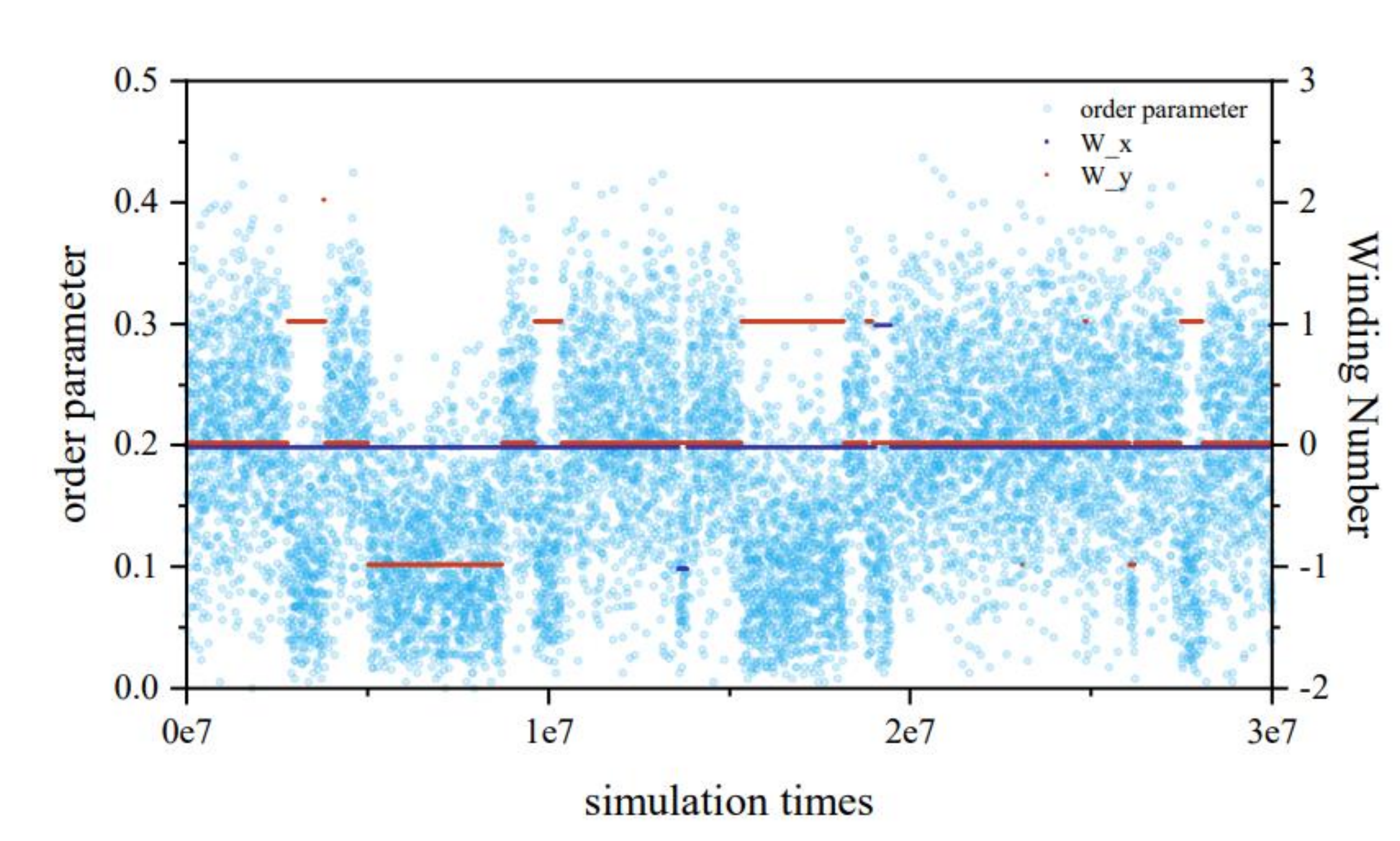}}
		\quad		
		\subfigure[]{ \label{7b} \includegraphics[scale=0.3]{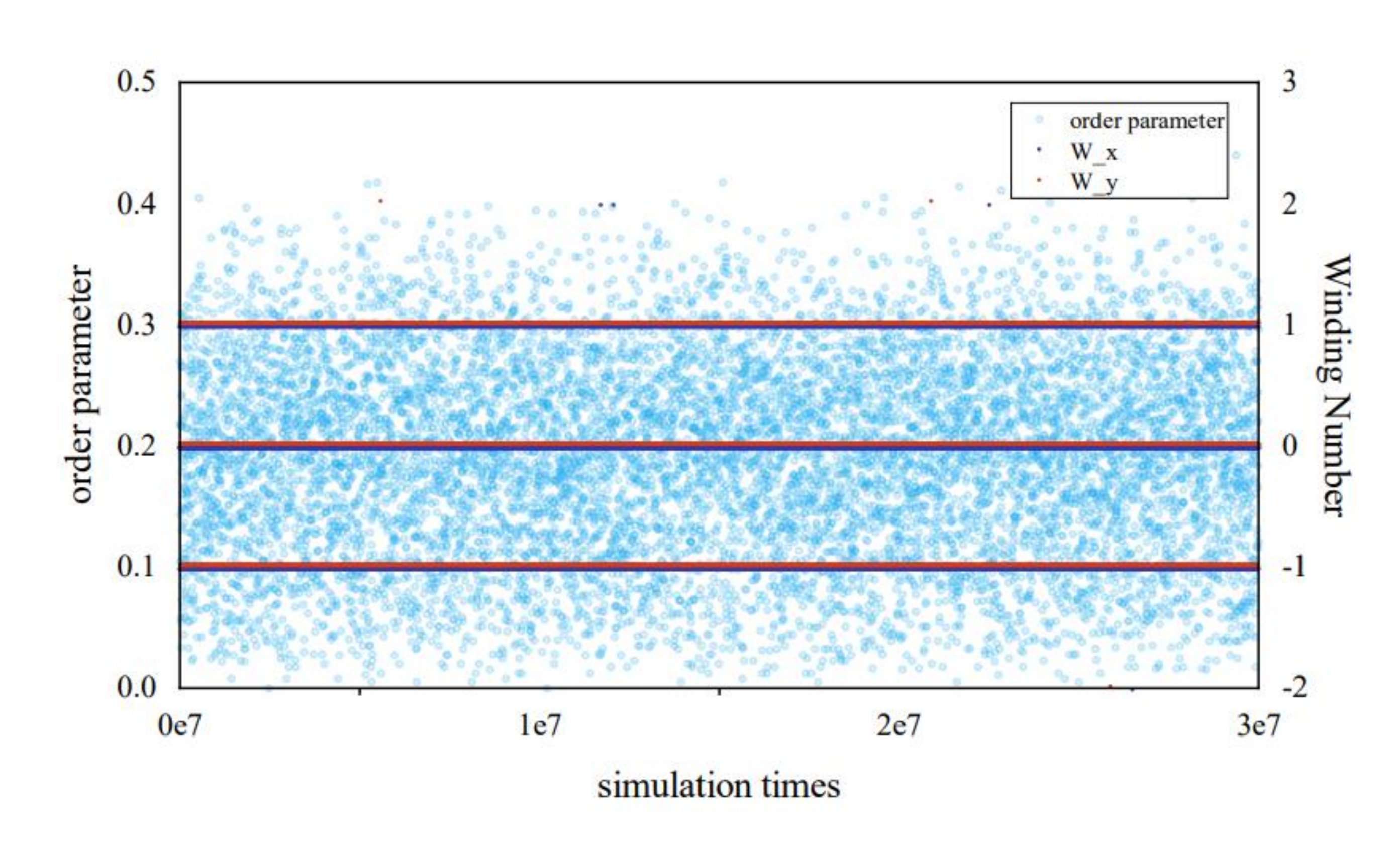}}
		\quad	
		\subfigure[]{ \label{7c} \includegraphics[scale=0.3]{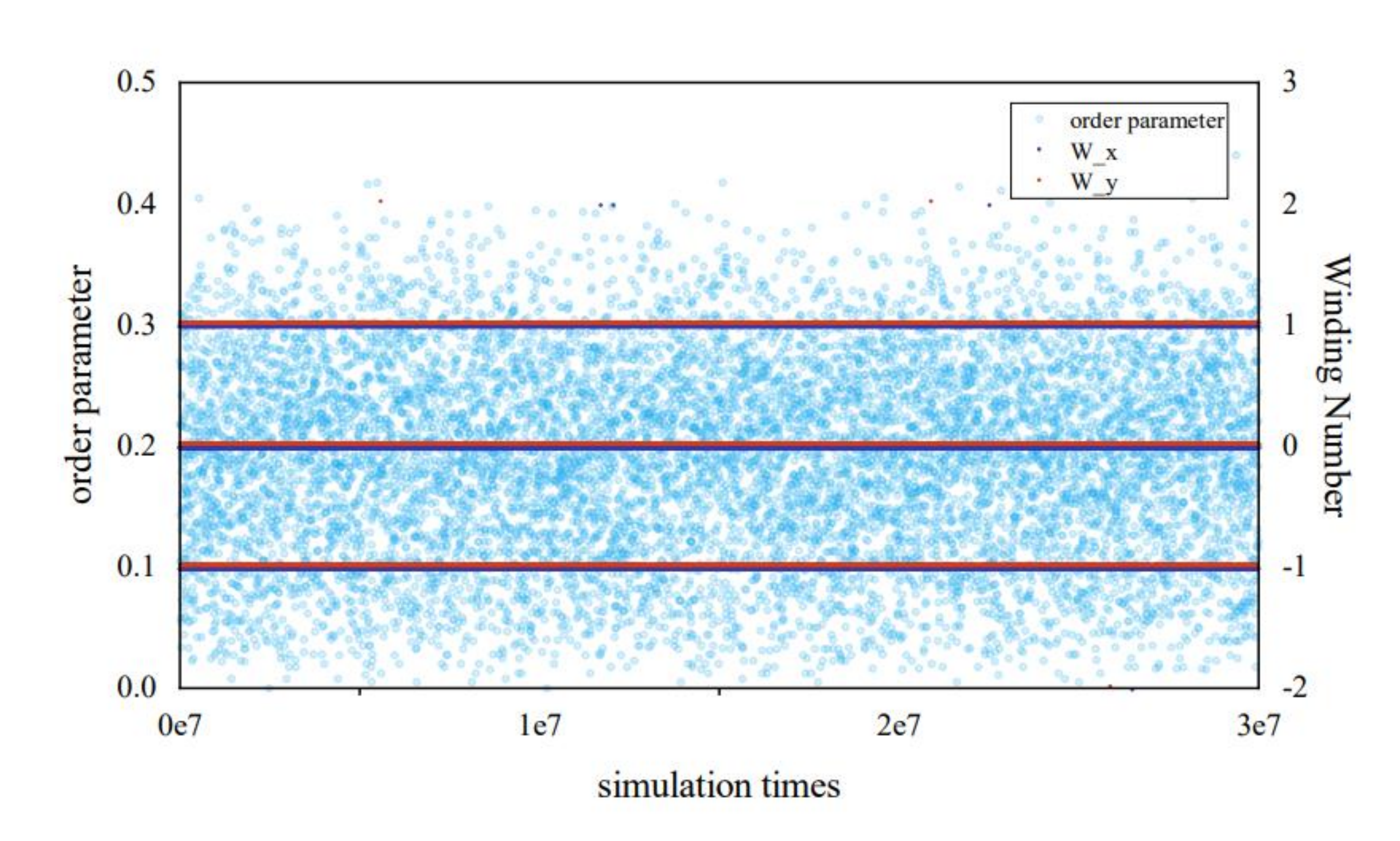}}
		\caption{(a)(b)(c): In 100,000 times simulation, evolvements along with simulation times of order parameter and topological winding number of DLA, eDLA and PEA.}		
	\end{figure}

	Strong autocorrelation will cause slowing down questions\cite{PhysRevLett.62.361,book}  in previous research, and in dimer model one key point corresponds to slowing-down questions is the blockade\cite{inbook} question in different topological sections. The whole ensemble of the microstates will be divided into different topological sections and every section has characteristic physical quantities in average. We measure order parameter and topological winding numbers synchronously as the simulation proceeds. We could see if we use DLA, order parameter will fluctuate as a lot of packets because DLA could hardly form a global loop to traverse topological sections and once the winding numbers has been changed, configurations will be locked in this section. That is the reason why slowing down questions happens in classical hard-core dimers. Both eDLA and PEA with 1 seed here can traverse topological sections rapidly with long global loops so that they eliminate slowing-down questions in dimer simulations. 

\section{CONCLUSIONS AND OUTLOOK}

	Although traditional DLA is a universal algorithm for theoretical strong correlated models, its efficiency constraints us computing more complex dimer models. Numeric results show our developed eDLA is provided with potential computational ability at finite temperatures. 
 
	In our comparisons, eDLA will possess fastest convergent speed from ground states to its characteristic states and smallest autocorrelations than other previous algorithms. Its essence of DLA gives it talent to be used in other SDMs\cite{yao2021breaking} and criterion endowed by us makes the loop generation process considering its temperature and energy. PEA as tricky algorithms can form bigger loops and multi-seeds added by us can improve its efficiency at high temperatures. But the specific generation of long loops will lead it least efficient among above algorithms. We try to develop PEA in SDMs but we can not fail to mention that the closure problem could decrease its efficiency. Both eDLA as a global algorithm and PEA as a non-local algorithm can traverse topological sections rapidly, so the simulation conclusions of Monte Carlo will be computed in a larger ensembles and derived more precisely. We have basically introduce credible methods in classical dimer model and other SDMs which gives us more chance to broaden the possibilities in dimer questions.

	\begin{acknowledgments}	

	\end{acknowledgments}

	\begin{figure*}[htbp]
		\centering
		\subfigure[]{ \label{8a} \includegraphics[scale=0.22]{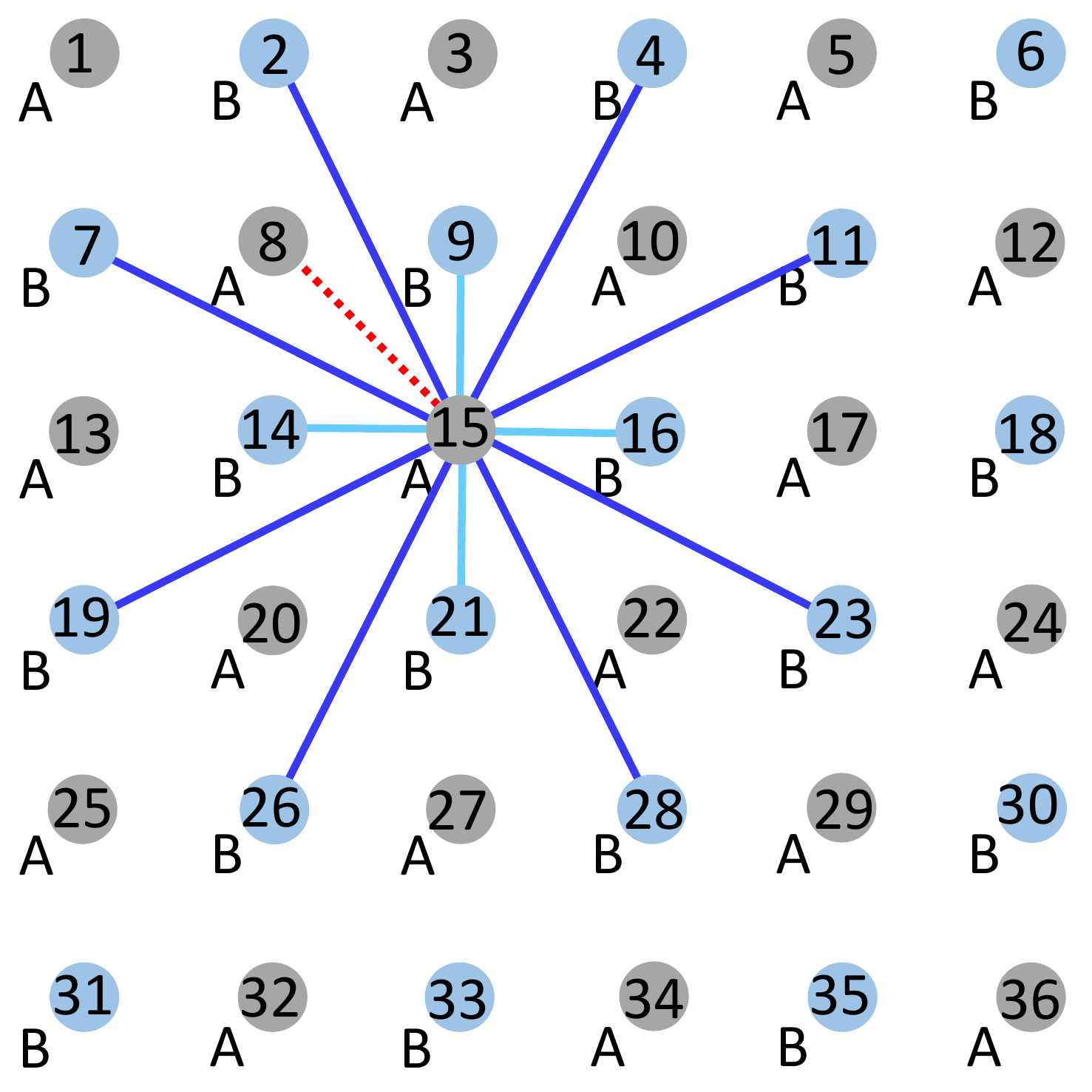}}
		\quad	
		\quad	
		\subfigure[]{ \label{8b} \includegraphics[scale=0.19]{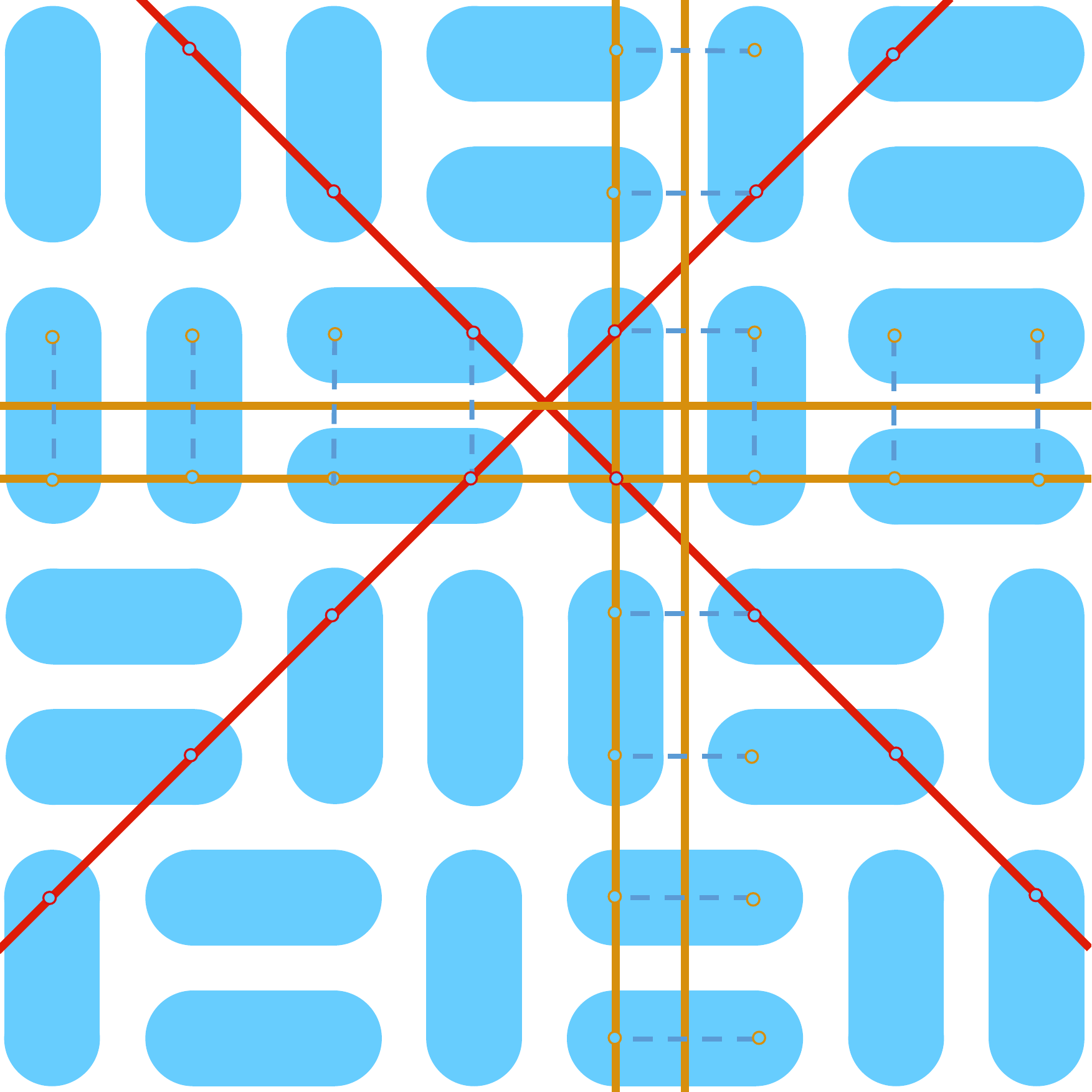}}
		\quad	
		\subfigure[]{ \label{8c} \includegraphics[scale=0.25]{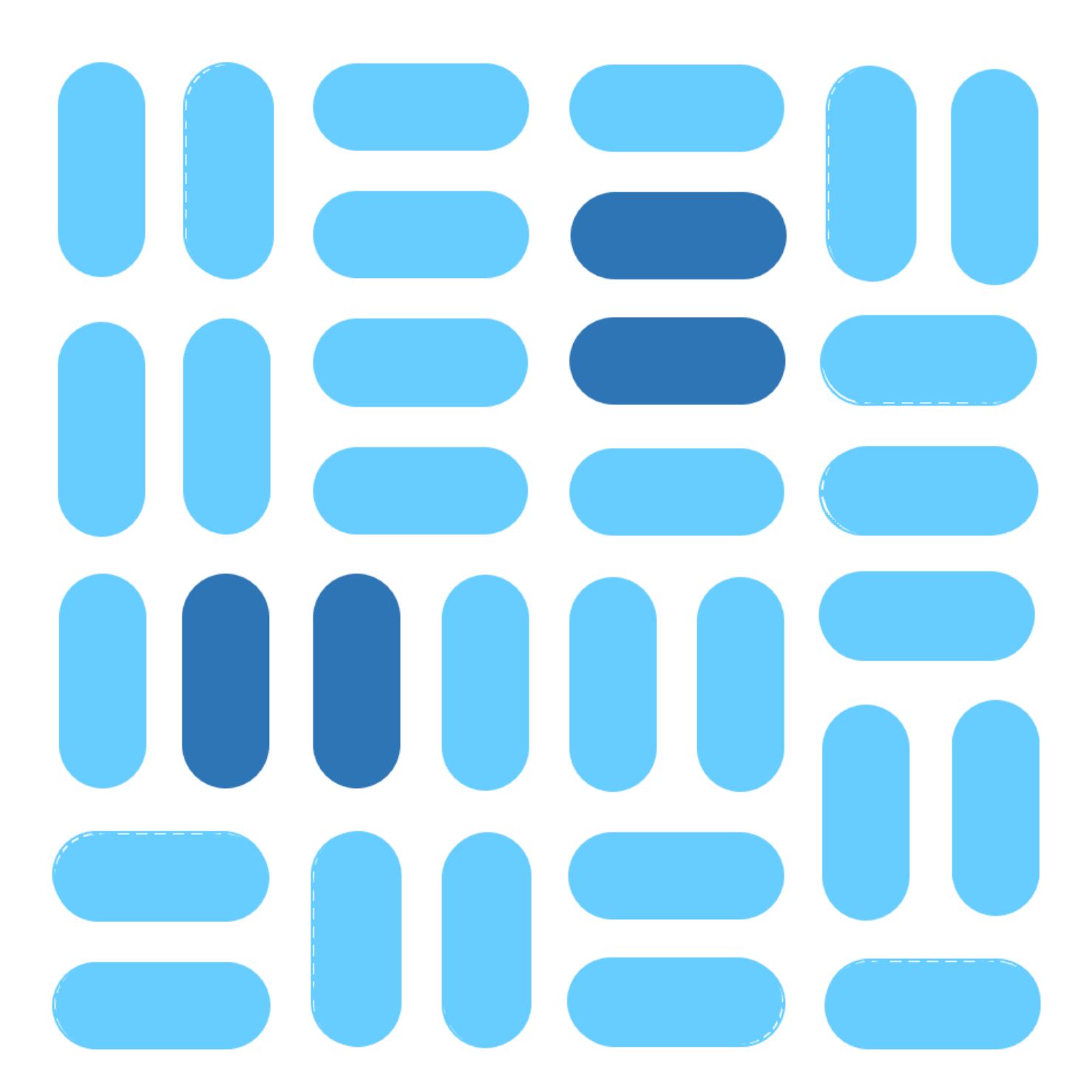}}
		\quad
		\subfigure[]{ \label{8d} \includegraphics[scale=0.25]{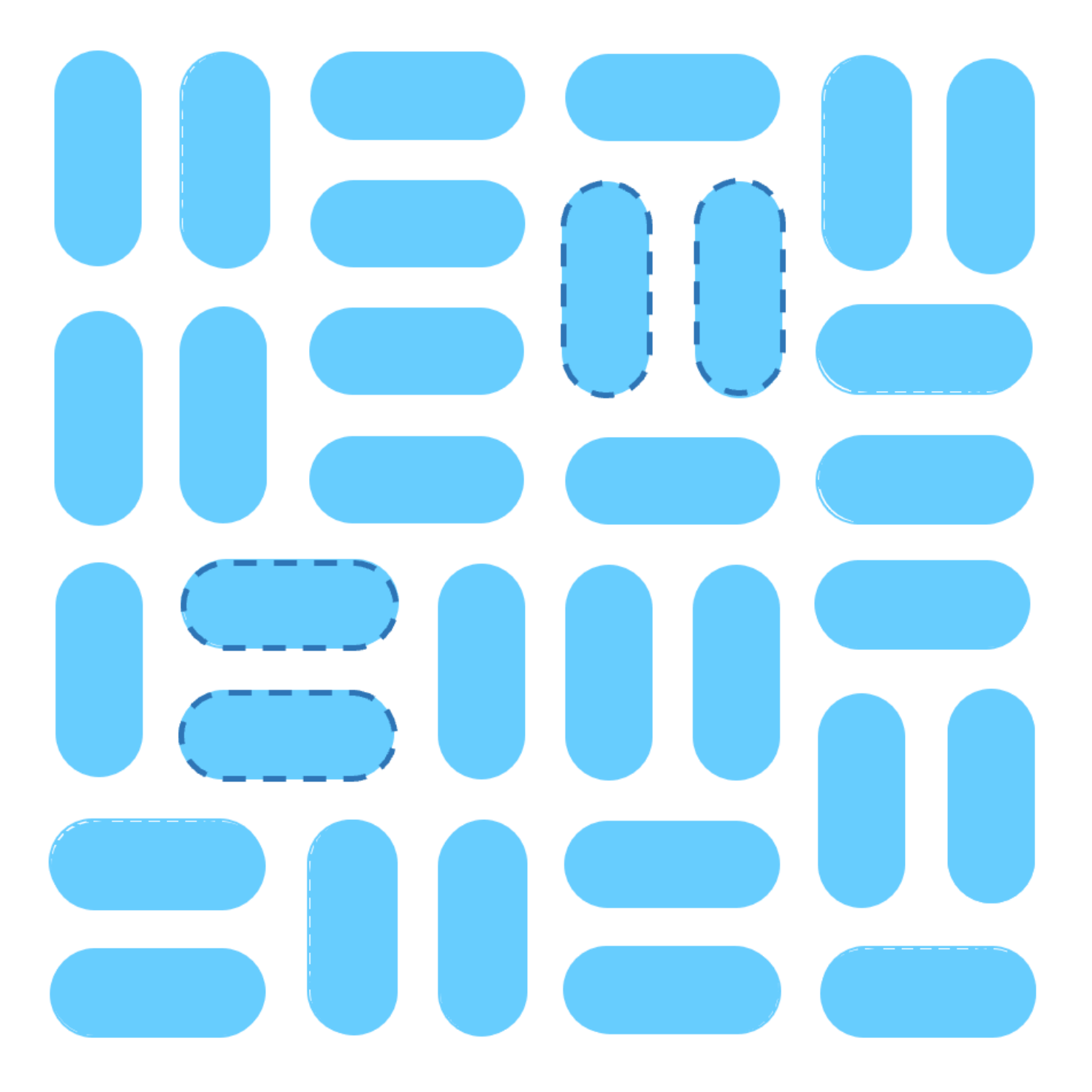}}
		\quad		
		\subfigure[]{ \label{8e} \includegraphics[scale=0.25]{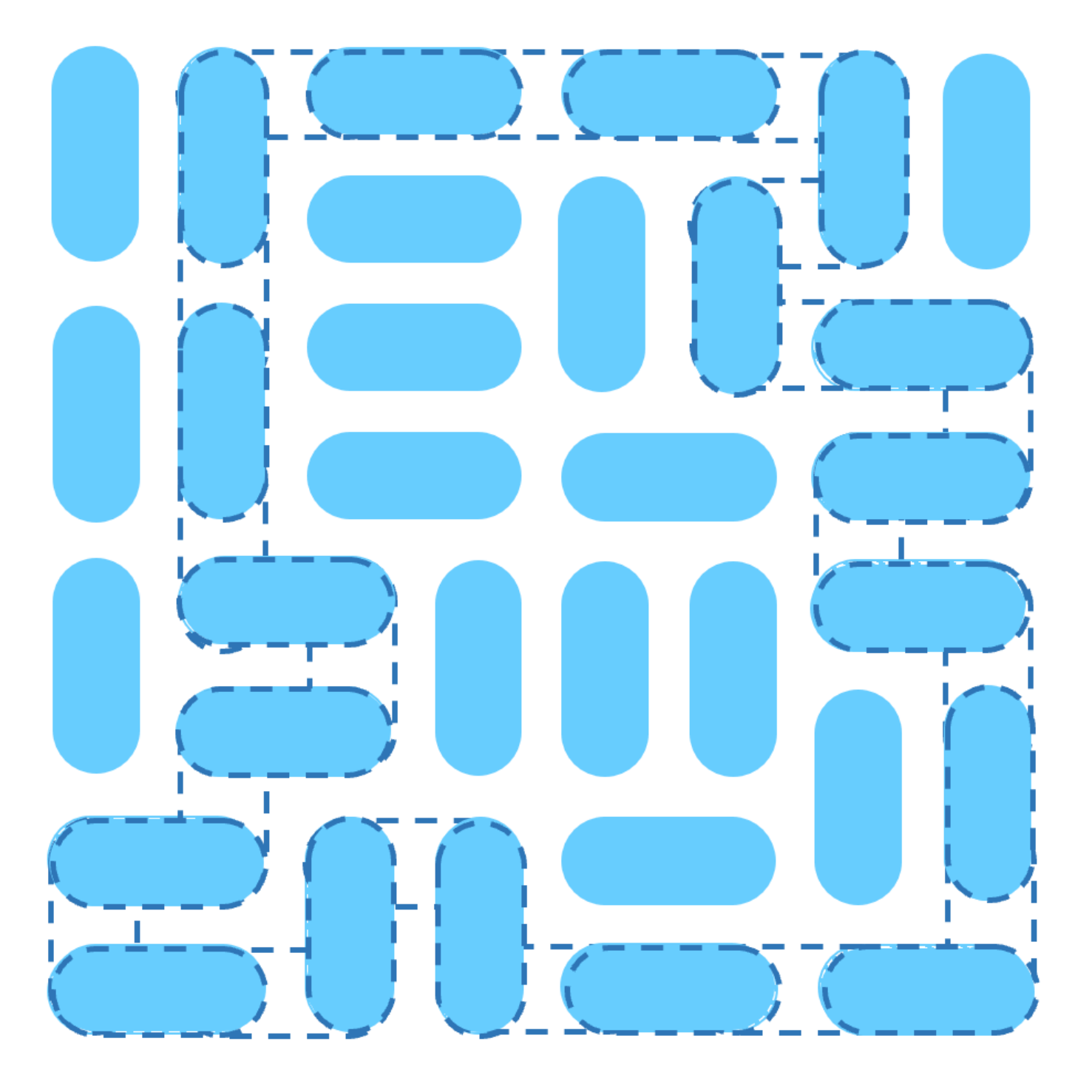}}
		\quad
		\subfigure[]{ \label{8f} \includegraphics[scale=0.25]{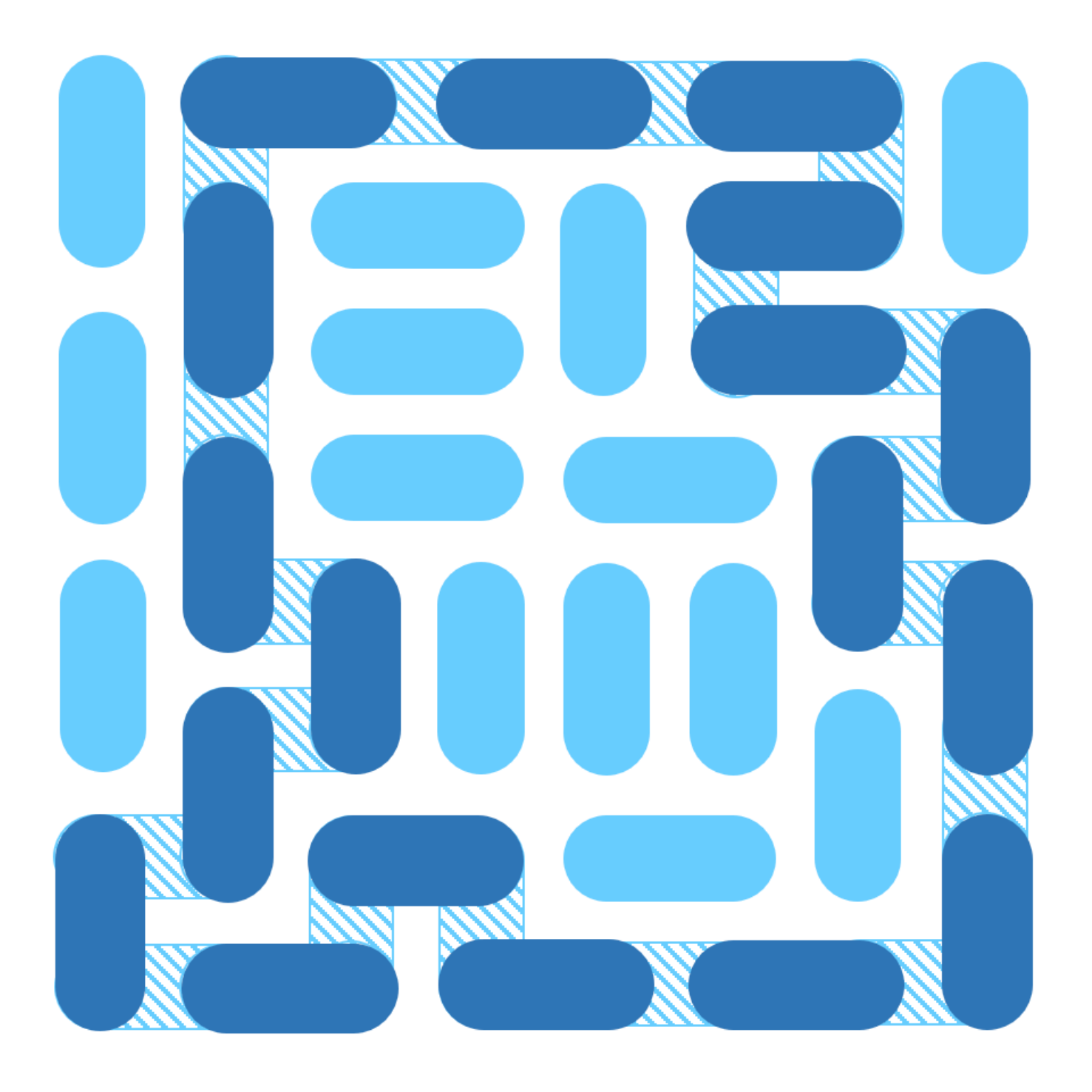}}
		\quad
		\subfigure[]{ \label{8g} \includegraphics[scale=0.4]{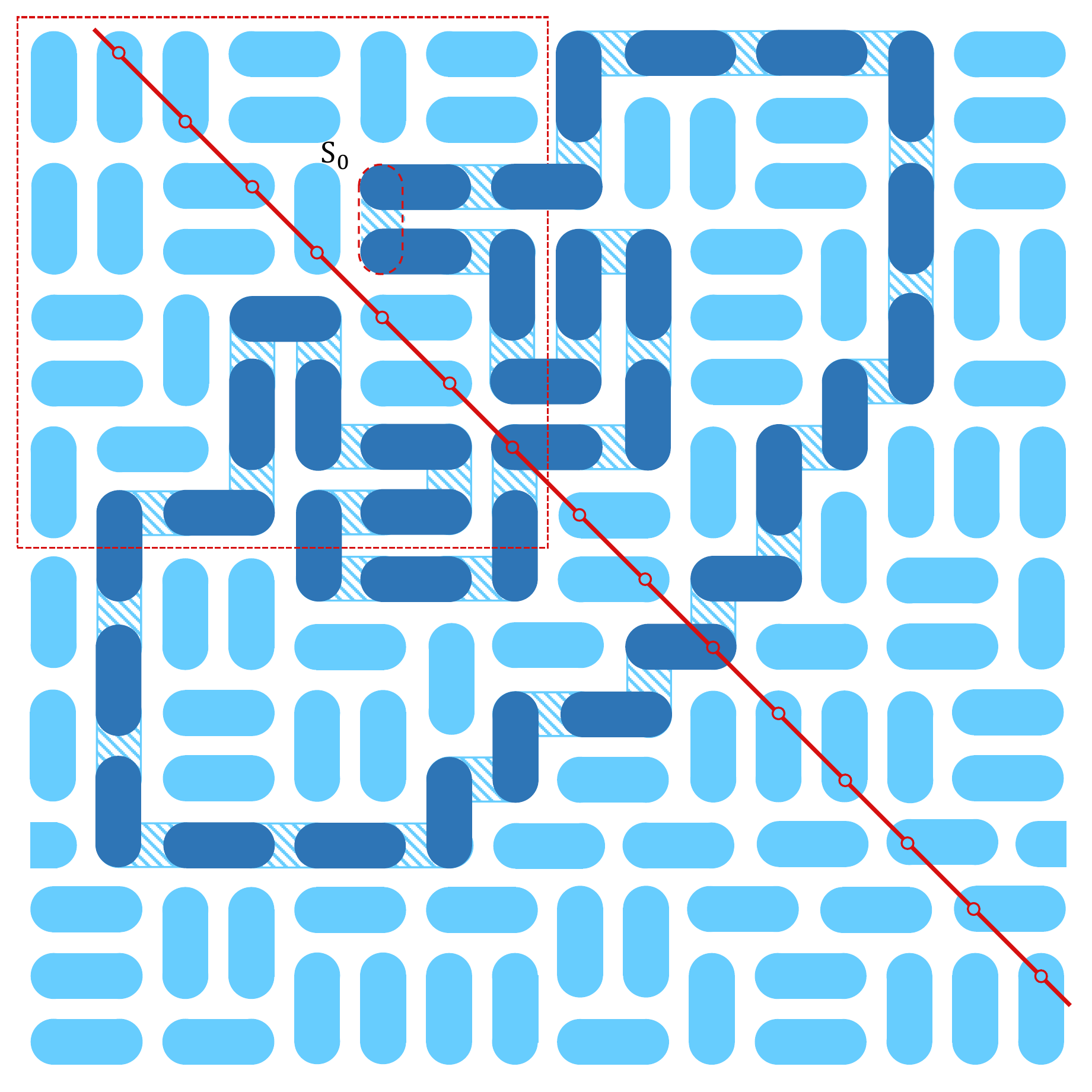}}
		\caption{(a): bipartite and non-bipartite lattice of dimers. Tint and deep blue lines are bipartite lattice bonding distinguishable vertex A and B. Tint blue lines are hard-core dimers and deep blue lines are N1-N4 bonds in SDM. Red dashed line is non-bipartite frustrated lattice bonding vertex A and A. (b): four symmetry axes used in PEA. (c)(d): Simplest local plaquette flappable algorithm. (deep blue dimers are ready to be flipped and dashed lines cover dimers flipped under an update.  (e)(f): traditional directed loop of  DLA (In Fig.8(e), dashed line is the directed loop. In Fig.8(f), dimers are be updated in turn along a directed loop marked in deep blue line). (g): the whole figure of Fig.1(e)(f). PEA can form a symmetry loop in the case of Fig.1(e)(f).}		
	\end{figure*}

\appendix
\section{DIRECTED LOOP ALGORITHM (DLA)}
	We compute dimer models to test our algorithms on bipartite square lattices and we can mark every vertex to define vector order parameter shown in Fig.8(a). Tint blue lines are hard-core dimers. If we consider deconfinement cases, rea dashed line is vertex bonded by next nearest points with frustrated condition ($N1-N2$ model) and deep blue lines are vertex bonded by second nearest points without frustrated condition ($N1-N4$ model) shown in Fig.8(b). Early researches use the DLA in the classical dimer model or SDMs broken geometric constraints\cite{doi:10.1063/1.1632141,2004Deconfinement,yao2021breaking}. For a given state in the Monte Carlo process, we can choose a point on the lattice randomly, it is used for the starting point of the loop and the forward of this dimer. The loop will pass through the dimer existed at this point spontaneously and there are 3 points near the bottom point of the chosen dimer (7 points if $N1-N2$ model or 11points if $N1-N4$ model). In three direction met by the bottom point, we choose one of them randomly and if we repeat the operation until the loop is closed see Fig.8(3)), we could shift all dimers to all pockets nearest to the dimer along the loop direction. Computing   the whole energy change and utilizing the Metropolis algorithm make us judge whether this update is accepted or not (see Fig.8(f)).

\bibliographystyle{unsrt}
\bibliography{Dimermodelarticle1}

\begin{thebibliography}{10}

\bibitem{PhysRevLett.61.2376}
Daniel~S. Rokhsar and Steven~A. Kivelson.
\newblock Superconductivity and the quantum hard-core dimer gas.
\newblock {\em Phys. Rev. Lett.}, 61:2376--2379, Nov 1988.

\bibitem{doi:10.1126/science.235.4793.1196}
P.~W. Anderson.
\newblock The resonating valence bond state in $la_2cuo_4$ and
  superconductivity.
\newblock {\em Science}, 235(4793):1196--1198, 1987.

\bibitem{doi:10.1080/14786439808206568}
P.~Fazekas and P.~W. Anderson.
\newblock On the ground state properties of the anisotropic triangular
  antiferromagnet.
\newblock {\em The Philosophical Magazine: A Journal of Theoretical
  Experimental and Applied Physics}, 30(2):423--440, 1974.

\bibitem{BASKARAN1987973}
G.~Baskaran, Z.~Zou, and P.W. Anderson.
\newblock The resonating valence bond state and high-tc superconductivity — a
  mean field theory.
\newblock {\em Solid State Communications}, 63(11):973--976, 1987.

\bibitem{PhysRevB.71.020401}
Olav~F. Sylju\aa{}sen.
\newblock Continuous-time diffusion monte carlo method applied to the quantum
  dimer model.
\newblock {\em Phys. Rev. B}, 71:020401, Jan 2005.

\bibitem{PhysRevB.73.245105}
Olav~F. Sylju\aa{}sen.
\newblock Plaquette phase of the square-lattice quantum dimer model: Quantum
  monte carlo calculations.
\newblock {\em Phys. Rev. B}, 73:245105, Jun 2006.

\bibitem{PhysRevLett.100.037201}
A.~Ralko, D.~Poilblanc, and R.~Moessner.
\newblock Generic mixed columnar-plaquette phases in rokhsar-kivelson models.
\newblock {\em Phys. Rev. Lett.}, 100:037201, Jan 2008.

\bibitem{PhysRevB.99.165135}
Zheng Yan, Yongzheng Wu, Chenrong Liu, Olav~F. Sylju\aa{}sen, Jie Lou, and Yan
  Chen.
\newblock Sweeping cluster algorithm for quantum spin systems with strong
  geometric restrictions.
\newblock {\em Phys. Rev. B}, 99:165135, Apr 2019.

\bibitem{PhysRevB.103.094421}
Zheng Yan, Zheng Zhou, Olav~F. Sylju\aa{}sen, Junhao Zhang, Tianzhong Yuan, Jie
  Lou, and Yan Chen.
\newblock Widely existing mixed phase structure of the quantum dimer model on a
  square lattice.
\newblock {\em Phys. Rev. B}, 103:094421, Mar 2021.

\bibitem{2021Topological}
Z~Yan, Yc~Wang, N~Ma, Y~Qi, and Z~Meng.
\newblock Topological phase transition and single/multi anyon dynamics of z2
  spin liquid.
\newblock {\em npj Quantum Mater}, 6.

\bibitem{doi:10.1063/1.1632141}
Anders~W. Sandvik and Olav~F. Syljuåsen.
\newblock The directed‐loop algorithm.
\newblock {\em AIP Conference Proceedings}, 690(1):299--308, 2003.

\bibitem{PhysRevE.71.036706}
Fabien Alet, Stefan Wessel, and Matthias Troyer.
\newblock Generalized directed loop method for quantum monte carlo simulations.
\newblock {\em Phys. Rev. E}, 71:036706, Mar 2005.

\bibitem{2004Deconfinement}
A.~W. Sandvik.
\newblock Deconfinement and criticality in extended two-dimensional dimer
  models, 2004.

\bibitem{PhysRevB.67.064503}
Werner Krauth and R.~Moessner.
\newblock Pocket monte carlo algorithm for classical doped dimer models.
\newblock {\em Phys. Rev. B}, 67:064503, Feb 2003.

\bibitem{2008LNP...739.....F}
H.~{Fehske}, R.~{Schneider}, and A.~{Wei{\ss}e}.
\newblock {\em {Computational Many-Particle Physics}}, volume 739.
\newblock 2008.

\bibitem{PhysRevLett.89.137202}
G.~Misguich, D.~Serban, and V.~Pasquier.
\newblock Quantum dimer model on the kagome lattice: Solvable dimer-liquid and
  ising gauge theory.
\newblock {\em Phys. Rev. Lett.}, 89:137202, Sep 2002.

\bibitem{PhysRevX.10.011005}
Felix Flicker, Steven~H. Simon, and S.~A. Parameswaran.
\newblock Classical dimers on penrose tilings.
\newblock {\em Phys. Rev. X}, 10:011005, Jan 2020.

\bibitem{PhysRevB.54.12938}
P.~W. Leung, K.~C. Chiu, and Karl~J. Runge.
\newblock Columnar dimer and plaquette resonating-valence-bond orders in the
  quantum dimer model.
\newblock {\em Phys. Rev. B}, 54:12938--12945, Nov 1996.

\bibitem{yan2021improved}
Zheng Yan.
\newblock Improved sweeping cluster algorithm for quantum dimer model, 2021.

\bibitem{yao2021breaking}
Hongxu Yao, Jiaze Li, and Jintao Hou.
\newblock The breaking of geometric constraint of classical dimers on the
  square lattice, 2021.

\bibitem{PhysRevLett.94.235702}
Fabien Alet, Jesper~Lykke Jacobsen, Gr\'egoire Misguich, Vincent Pasquier,
  Fr\'ed\'eric Mila, and Matthias Troyer.
\newblock Interacting classical dimers on the square lattice.
\newblock {\em Phys. Rev. Lett.}, 94:235702, Jun 2005.

\bibitem{KITAEV20032}
A.Yu. Kitaev.
\newblock Fault-tolerant quantum computation by anyons.
\newblock {\em Annals of Physics}, 303(1):2--30, 2003.

\bibitem{RN18}
Ernst Ising.
\newblock Beitrag zur theorie des ferromagnetismus.
\newblock {\em Zeitschrift für Physik}, 31(1):253--258, 1925.

\bibitem{PhysRevLett.62.361}
Ulli Wolff.
\newblock Collective monte carlo updating for spin systems.
\newblock {\em Phys. Rev. Lett.}, 62:361--364, Jan 1989.

\bibitem{book}
Burkhard Dünweg, D.~Landau, and Andrey Milchev.
\newblock {\em Computer Simulations of Surfaces and Interfaces}.
\newblock 01 2003.

\bibitem{inbook}
Daan Frenkel and Berend Smit.
\newblock {\em Accelerating Monte Carlo Sampling}, pages 399--403.
\newblock 01 2002.

\end{thebibliography}
\end{document}